\documentclass[a4paper,USenglish,cleveref,thm-restate]{lipics-v2021}

\pdfoutput=1 
\hideLIPIcs  

\usepackage{bm}
\usepackage{xspace}
\usepackage[noend]{algpseudocode}

\nolinenumbers
\title{Almost Optimal Algorithms for Token Collision\\ in Anonymous Networks}

\titlerunning{Almost Optimal Algorithms for Token Collision in Anonymous Networks}

\author{Sirui Bai}{State Key Laboratory for Novel Software Technology, Nanjing University, China}{srbai@smail.nju.edu.cn}{}{}

\author{Xinyu Fu}{State Key Laboratory for Novel Software Technology, Nanjing University, China}{xyfu@smail.nju.edu.cn}{}{}

\author{Xudong Wu}{State Key Laboratory for Novel Software Technology, Nanjing University, China}{xdwu@smail.nju.edu.cn}{}{}

\author{Penghui Yao}{State Key Laboratory for Novel Software Technology, Nanjing University, China\\Hefei National Laboratory, Hefei 230088, China}{pyao@nju.edu.cn}{}{}

\author{Chaodong Zheng}{State Key Laboratory for Novel Software Technology, Nanjing University, China}{chaodong@nju.edu.cn}{}{}

\authorrunning{S. Bai, X. Fu, X. Wu, P. Yao, and C. Zheng}

\Copyright{Sirui Bai, Xinyu Fu, Xudong Wu, Penghui Yao, and Chaodong Zheng}



\category{} 

\relatedversion{} 

\funding{This work was supported by National Natural Science Foundation of China (Grant No. 62332009, 62172207, 12347104), the Department of Science and Technology of Jiangsu Province of China (Grant No. BK20211148) and Innovation Program for Quantum Science and Technology (Grant No. 2021ZD0302901).} 

\EventEditors{Dan Alistarh}
\EventNoEds{1}
\EventLongTitle{38th International Symposium on Distributed Computing (DISC 2024)}
\EventShortTitle{DISC 2024}
\EventAcronym{DISC}
\EventYear{2024}
\EventDate{October 28--November 1, 2024}
\EventLocation{Madrid, Spain}
\EventLogo{}
\SeriesVolume{319}
\ArticleNo{33}

\newcommand{\highlightblue}[1]{{\color{black} #1}}
\newcommand{\highlight}[2][black]{{\color{#1} #2}}

\newtheorem{fact}[theorem]{Fact}

\makeatletter
\algnewcommand{\LineComment}[1]{\Statex\hskip\ALG@thistlm $\triangleright$ #1}
\makeatother

\DeclareMathOperator*{\argmin}{argmin}

\newcommand{\vid}{\texttt{id}}
\newcommand{\rid}{\texttt{rid}}
\newcommand{\parent}{\texttt{p}}
\newcommand{\ischild}{\texttt{ischild}}
\newcommand{\child}{\texttt{chi}}
\newcommand{\flagnew}{\texttt{f}}
\newcommand{\cnt}{\texttt{cnt}}
\newcommand{\build}{\texttt{build}}
\newcommand{\res}{\texttt{res}}
\newcommand{\ele}{\texttt{ele}}

\newcommand{\sent}{\texttt{sent}}
\newcommand{\sente}{\texttt{sente}}
\newcommand{\pieces}{M}
\newcommand{\eachpiece}{P}
\newcommand{\rounds}{Q}
\newcommand{\ceillb}{{\lceil L/B\rceil}}

\newcommand{\dist}{\texttt{dist}}

\begin{document}

\maketitle

\begin{abstract}
\highlightblue{
In distributed systems, situations often arise where some nodes each holds a collection of \emph{tokens}, and all nodes collectively need to determine whether all tokens are distinct. For example, if each token represents a logged-in user, the problem corresponds to checking whether there are duplicate logins. Similarly, if each token represents a data object or a timestamp, the problem corresponds to checking whether there are conflicting operations in distributed databases. In distributed computing theory, unique identifiers generation is also related to this problem: each node generates one token, which is its identifier, then a verification phase is needed to ensure all identifiers are unique.

In this paper, we formalize and initiate the study of \emph{token collision}. In this problem, a collection of $k$ tokens, each represented by some length-$L$ bit string, are distributed to $n$ nodes of an {\em anonymous} CONGEST network in an arbitrary manner. The nodes need to determine whether there are tokens with an identical value. We present near optimal deterministic algorithms for the token collision problem with $\tilde{O}(D+k\cdot L/\log{n})$ round complexity, where $D$ denotes the network diameter. Besides high efficiency, the prior knowledge required by our algorithms is also limited. For completeness, we further present a near optimal randomized algorithm for token collision.
}
\end{abstract}


\section{Introduction}\label{sec:intro}

Imagine the following scenario: a group of servers is hosting an online-banking, online-gaming, or online-exam service; for security reasons, users are not allowed to log into multiple servers simultaneously. If we interpret each logged-in user as a \emph{token}, the servers need to check whether all active tokens are distinct. Similar problems could also arise in distributed database systems. For example, in some distributed databases, optimistic concurrency control schemes are employed to increase concurrency and performance~\cite{kung81}. Motivation for such schemes is the observation that system clients are unlikely to access the same object concurrently. Nonetheless, before the system commit clients' transactions, verification must be performed to ensure that all read and write operations are disjoint or that no operations occur at the same time, otherwise a rollback is necessary. In this setting, tokens represent data objects or timestamps~\cite{neumann15,ozsu20}.

Apart from practical scenarios, detecting colliding tokens is also important from a theoretical perspective as we can interpret identifiers as tokens. Specifically, it is well-known that a number of fundamental distributed computing tasks, such as coloring, leader election, bipartiteness testing, and planarity testing, are impossible to resolve deterministically in anonymous networks~\cite{angluin80,johnson85,peleg00}. Hence, unique identifiers generation becomes an important primitive for anonymous networks, as it breaks the symmetry within the network, thus making the above tasks possible.
Moreover, the lengths of these identifiers could affect the performance of corresponding algorithms; examples include renaming algorithms, Linial's classical $n$-to-$\Delta^2$ coloring algorithm~\cite{peleg00}, recent deterministic network decomposition algorithms by Ghaffari et al.~\cite{ghaffari21}, and recent MST algorithm focusing on energy complexity~\cite{augustine22}.
A plausible approach for generating unique identifiers in anonymous networks is to employ randomness: for instance, each of the $n$ nodes generates an identifier by sampling $\Theta(\log{n})$ uniform random bits; by the birthday paradox, all identifiers are distinct with probability at least $1-1/n$. However, this is a Monte Carlo algorithm that is subject to error. If we seek a Las Vegas (i.e., zero-error) algorithm for generating unique identifiers in anonymous networks, a deterministic algorithm for detecting colliding identifiers (i.e., tokens) is necessary: nodes repeatedly run a (Monte Carlo) randomized identifiers generation algorithm and use a deterministic algorithm to check whether the generated identifiers are unique.

Despite the various applications for \emph{token collision}, somewhat surprisingly, this problem has not been explicitly studied in the context of distributed computing to the best of our knowledge. In this paper, we initiate the study of this generic distributed computing task, and we begin by giving a definition for it:

\highlightblue{
\begin{definition}[\textbf{Token Collision}]\label{def:problem}
Assume that there are $k$ tokens, each having a value represented by a length-$L$ binary string. Consider a distributed system consisting of $n$ nodes. The $k$ tokens are divided into $n$ collections, some of which may be empty. Each of the $n$ nodes is assigned one collection as input. In the \emph{token collision} problem, the nodes need to determine whether there are tokens with identical value.
\end{definition}
}

We focus on understanding the time complexity of token collision in anonymous networks. The reason for considering anonymous networks is two-fold.
First, if nodes in a distributed system already have unique identifiers, then token collision (or almost any distributed computing task) could be resolved by first electing the node with the smallest identifier as the leader, then aggregate necessary information to the leader, and finally let the leader locally compute the result and disseminate the result to the rest of the network.\footnote{Nonetheless, our lower bounds for token collision hold even for named networks, and our algorithms nearly match these lower bounds. Thus, anonymity does not make token collision harder.}
Second, anonymous networks also arise in real-world scenarios. For example, the nodes in a distributed system (e.g., sensor networks) may be indistinguishable since they are fabricated in a large-scale industrial process, in which equipping every node with a unique identifier is not economically feasible (e.g., MAC addresses are not necessarily unique nowadays). In other cases nodes may not wish to reveal their identities out of privacy or security concerns.

We consider standard CONGEST model~\cite{peleg00} in distributed computing. A CONGEST network is described by a graph $G=(V,E)$ with $|V|=n$ nodes being processors with unlimited computational power (we do not exploit this ability in this paper) and edges being communication channels with bounded bandwidth. Specifically, we assume that any message sent through a channel cannot exceed $\Theta(\log{n})$ bits. \highlightblue{To simplify presentation, we often use $B=\Theta(\log{n})$ to denote this bandwidth limitation.} Processors exchange messages synchronously, round-by-round along the channels. When proving impossibility results for the token collision problem, we also consider an alternate model known as the LOCAL model~\cite{linial92}, where communication channels have unbounded bandwidth.

\subsection{Results and contribution}

\subparagraph{Deterministic scenario.} We first consider the case where the tokens are not too large---particularly, every token can fit into one message. In this scenario, we offer a deterministic algorithm that works so long as every node knows the exact value of $n$ or $k$.

\highlightblue{
\begin{theorem}[\textbf{Deterministic Upper Bound, Part 1}]\label{thm:upper-det-part1}
In an $n$-node anonymous CONGEST network with diameter $D$, for any instance of the token collision problem in which $k$ tokens are encoded by length-$L$ bit strings, if every node knows the exact value of $n$ or $k$, then there exists an $O(D+k\cdot L/\log n)$-round deterministic algorithm when $L=O(\log n)$.
\end{theorem}
}

The above theorem implies when $L=\Theta(\log{n})$, token collision can be resolved within $O(D+k)$ rounds. As a result, one could easily derive a Las Vegas unique identifiers generation algorithm with $O(n)$ expected runtime. On the other hand, for problem instances where tokens are small---$L=o(\log{n})$ in particular, the runtime of the algorithm is $O(D)+o(k)$.


At a high level, our algorithm tries to find the token(s) that have the global minimum value and selects the node(s) that own(s) such token(s) as leader(s). Critically, our algorithm may elect multiple nodes as leaders, so it does not solve the leader election problem. (In fact, leader election cannot be solved deterministically in our setting.) Nevertheless, alongside this election process, BFS-trees will be built with these leaders being the roots, thus the network graph becomes a forest logically. Then, by convergecasting~\cite{welch04} tokens within each tree and computing the size of each tree, root nodes can correctly determine the result. Similar ideas have been used in the design of Las Vegas leader election algorithms, \highlightblue{but the analysis of our deterministic algorithm is more challenging, see \Cref{subsec:related-work} for more discussion.}

We then extend our algorithm to the scenario where each token cannot fit into one message, this could occur in applications like plagiarism checking in which each token is a text segment. In this case, a simple solution is to divide each token into $\Theta(L/\log{n})$ parts, and use multiple rounds to simulate one round of our above algorithm. However, the resulting algorithm would have a round complexity of \highlightblue{$O((D+k)\cdot L/\log n)$}, which is too large. Instead, we devise a variant that uses pipelining techniques and extend the analysis accordingly. The following theorem states the time complexity of this variant.

\highlightblue{
\begin{theorem}[\textbf{Deterministic Upper Bound, Part 2}]\label{thm:upper-det-part2}
In an $n$-node anonymous CONGEST network with diameter $D$, for any instance of the token collision problem in which $k$ tokens are encoded by length-$L$ bit strings, if every node knows the exact value of $n$ or $k$, then there exists an \highlight{$O(D\cdot\max\{(\log(L/\log n))/\log n,1\}+k\cdot L/\log n)$}-round deterministic algorithm when $L=\omega(\log n)$.
\end{theorem}
}

To complement the algorithmic results, we have also established a lower bound on the round complexity of the token collision problem.

\begin{theorem}[\textbf{Deterministic Lower Bound}]\label{thm:lower-det}
In an $n$-node anonymous \highlightblue{CONGEST} network with diameter $D$, there are instances of the token collision problem in which $k$ tokens are encoded by length-$L$ bit strings such that, any deterministic algorithm requires \highlight{$\Omega(D+k\cdot (L-\log{k}+1)/\log n)$} rounds to solve it when $2^L\ge k$.
\end{theorem}

It is easy to verify, if $L\geq(1+\delta)\log{k}$ for some constant $\delta>0$, then the round complexity of our algorithm is tight when \highlightblue{$L=O(\log n)$}, and near-optimal (within multiplicative \highlightblue{$\log(L/\log n)/\log n$} factor) when \highlightblue{$L=\omega(\log n)$}.
\highlightblue{We also note that the assumption of $2^L\geq k$ is without loss of generality, as otherwise collision must occur.}
To obtain the lower bound, we reduce a variant of the set-disjointness problem in the study of two-party communication complexity to the token collision problem.

Another advantage of our algorithm is that it requires little prior knowledge. Beside input tokens, each node only needs to know the value of $n$ \emph{or} the value of $k$. In particular, nodes do not need to know the network diameter $D$. In fact, we can prove via an indistinguishability argument that without any global knowledge, deterministic token collision detection is impossible.
\highlightblue{(Nonetheless, what is the minimal prior knowledge required remains to be an interesting open question.)}

\begin{theorem}[\textbf{Impossibility Result}]\label{thm:impossible}
In the anonymous LOCAL model, if every node has no knowledge regarding the network graph except being able to count and communicate over adjacent links locally, and if every node also has no knowledge regarding the tokens except the ones given as local input, then there is no deterministic algorithm that solves the token collision problem.
\end{theorem}

\subparagraph{Randomized scenario.} We also investigate the randomized round complexity of token collision. At the upper bound side, selecting a unique leader can be easily achieved with desirable probability when randomness is allowed. Then a random hash function is employed to reduce the bit-length of tokens if they are too large, without increasing the probability of collision significantly in case tokens are distinct. Finally, with a convergecast process similar to the deterministic algorithm, the unique leader can collect all tokens and determine whether there are collisions. At the lower bound side, we again employ the strategy of reduction, and utilize existing results on the hardness of randomized set-disjointness to obtain the desired result. Our findings for the randomized scenario are summarized below. Note that in contrast with the deterministic setting, the length of the token $L$ no longer appears in the lower bound, and $L$'s impact on the upper bound is also limited.

\begin{theorem}[\textbf{Randomized Upper and Lower Bound}]\label{thm:rnd-upper-lower}
Consider an $n$-node anonymous \highlightblue{CONGEST} network with diameter $D$. For any instance of the token collision problem with $k$ tokens, if every node knows the exact value of $n$ or $k$, then there exists a randomized algorithm that solves it in \highlight{$O(D\cdot\max\{(\log(\log{k}/\log n))/\log n,1\}+k\cdot\log{k}/\log n+L/\log n)$} rounds \highlightblue{with probability at least $1-1/k$}. On the other hand, assuming $2^L\geq k$, there are instances of the token collision problem with $k$ tokens such that any randomized algorithm that succeeds with probability at least $2/3$ requires \highlightblue{$\Omega(D+k/\log n)$} rounds.
\end{theorem}

\subsection{Related work and discussion}\label{subsec:related-work}

Though token collision has not been explicitly studied in distributed computing, similar problems have been investigated elsewhere.
For example, element distinctness, which decides whether a given set of elements are distinct, has been extensively studied in the context of query complexity.
Specifically, linear lower bounds were proved for deterministic and randomized algorithms~\cite{ben-or83,grigoriev96}.
A sublinear quantum algorithm was proposed by Buhrman et al.~\cite{buhrman05}, which applies $O(n^{3/4})$ quantum queries.
The upper bound was later improved to $O(n^{2/3})$ by Ambainis using quantum walk~\cite{ambainis07}, and matched the lower bound given by Aaronson and Shi~\cite{aaronson04}.
To the best of our knowledge, this paper is the first one that studies token collision in classical distributed computing models, focusing on round complexity.

Token collision is also related to leader election, a classical and fundamental distributed computing primitive, in several aspects.

On the one hand, the design of Las Vegas leader election algorithms~\cite{itai90,tel94,fokkink04} share similar ideas with ours. In those algorithms, usually nodes first randomly generate identifiers, then the node with the smallest identifier is elected as the leader if that identifier is owned by a single node, otherwise the process restarts. As can be seen, the problem of checking whether the smallest identifier is unique is a variant of the token collision problem. Indeed, the routine developed by Tel in \cite{tel94} for this checking procedure is very similar to  our algorithm. Nonetheless, the analysis in our setting is more involved: in the context of Las Vegas leader election, restart when the smallest identifier is unique is fine (i.e., false negative is fine), yet in our context this is unacceptable. In fact, proving such false negative will not occur is highly non-trivial (see \Cref{lem:tree-of-dinstinct} in \Cref{sec:alg-analysis}).
Moreover, our algorithm can handle the scenario that tokens are of arbitrary size, making it more generic.

On the other hand, as mentioned earlier, with a unique leader almost any distributed computing problem can be solved. This observation raises the question that whether defining and studying token collision is necessary. We believe the answer is positive. First, the leader election approach is not necessarily better. Taking the unique identifiers generation problem as an example, the approach of ``first elect a leader and then let the leader aggregate and check whether the generated identifiers are unique'' share same round complexity with our algorithm. Second, and more importantly, in situations where leader election is infeasible (e.g., deterministic leader election in anonymous networks, \highlight{randomized leader election that always terminates in anonymous rings of unknown size})~\cite{tel00}, our algorithm still works with deterministic correctness and time complexity guarantees.

\section{Preliminary}\label{sec:preliminary}

In this section, we briefly introduce some known results on the communication complexity of the set-disjointness problem as it is used in our lower bound proof.

Communication complexity was introduced by Yao~\cite{yao79}, which is nowadays a versatile method to prove lower bounds in distributed computing.
In the two-party communication complexity model, two players Alice and Bob, respectively, receive $x\in\mathcal X$ and $y\in\mathcal Y$ as input and need to compute $f(x,y)$, where $f:\mathcal X\times\mathcal Y\to\mathcal Z$ is a two-argument function.
The communication complexity of $f$ is the minimum number of bits Alice and Bob need to exchange to compute $f(x,y)$ for any input $x$ and $y$.

The problem of set-disjointness (denoted as ${\rm DISJ}$) is one of the most well-studied problems in communication complexity~\cite{kalyanasundaram92,razborov92,bar-yossef04}, where Alice and Bob are given a set, respectively, and they need to decide whether their sets are disjoint.
In this paper, we are interested in a variant of set-disjointness: ${\rm DISJ}^p_q:\tbinom{[p]}q\times\tbinom{[p]}q\to\{0,1\}$.
Alice is given a set $S\subseteq[p]$ and Bob is given a set $T\subseteq[p]$, where $|S|=|T|=q$.
They aim to determine whether the two sets are disjoint, that is, ${\rm DISJ}^p_q(S,T)=1$ iff $S\cap T=\emptyset$.
The communication complexity of ${\rm DISJ}^p_q$ is established by H{\aa}stad and Widgerson~\cite{hastad07}.

\begin{fact}[\cite{hastad07}]\label{fact:disj}
For every $q\le p/2$, $D({\rm DISJ}^p_q)=\Omega(\log\tbinom pq)$ and $R_{1/3}({\rm DISJ}^p_q)=\Omega(q)$.
Here, $D({\rm DISJ}^p_q)$ denotes the deterministic communication complexity of problem ${\rm DISJ}^p_q$, and $R_{1/3}({\rm DISJ}^p_q)$ denotes the randomized communication complexity of problem ${\rm DISJ}^p_q$ with the probability of error being at most $1/3$.
\end{fact}

\section{The Deterministic Algorithm}\label{sec:alg-det}

In this section, we focus on the most common scenario where each token can be fitted into one message \highlightblue{(that is, $L=O(\log n)$)}. We will extend our algorithm to other settings later.

Broadly speaking, our algorithm can be divided into two parts: the first part concerns with building rooted BFS-trees, while the second part concerns with calculating the size of the BFS-trees and aggregating tokens at the roots for decision-making.
Although the high-level idea of our algorithm is not complicated, implementing it correctly and efficiently is non-trivial, especially in the setting where nodes only have limited global knowledge.

We now describe the algorithm in detail. (Complete pseudocode of the algorithm is provided in \Cref{sec-app:pseudocode}.)

\subparagraph{Build BFS-tree(s).} Initially, each node $v$ sets its identifier to be the smallest token it received as input, or a special symbol if $v$ received no token. Then, it attempts to construct a BFS-tree rooted at itself by broadcasting its identifier to its neighbors in each round. Whenever $v$ receives a smaller identifier from some neighbor $u$, it updates its identifier to match that of neighbor $u$. Moreover, it designates $u$ as its parent and sends a notification to its parent in all subsequent rounds. As a result, whenever $v$ changes its identifier to that of some neighbor $u$'s, node $v$ is appending the BFS-tree rooted at itself to the BFS-tree that includes $u$. We note that each node $v$ uses a variable $\texttt{rid}_v$ to store its identifier. Intuitively, $\texttt{rid}_v$ stores the root's identifier of the BFS-tree that $v$ belongs to.\footnote{This is merely an ``intuition'' and not always true during BFS-tree construction, as identifiers are propagating gradually and tree shape may change frequently.} We also note that each node $v$ uses an integer $\texttt{p}_v\in[\Delta_v]$ to store its parent, where $\Delta_v$ is the degree of $v$. That is, each node $v$ locally labels each incident edge with a unique integer in $[\Delta_v]$, and uses the edge label as its local identity for the node at the other endpoint of the edge.

When node $v$ discovers that all of its neighbors share the same identifier as itself, it attempts to ascertain whether the BFS-tree rooted at itself is fully constructed. To this end, note that the BFS-tree rooted at node $v$ consists of node $v$ and the BFS-trees rooted at its children. Therefore, our algorithm's criterion for node $v$ to confirm that the BFS-tree rooted at itself is fully constructed is: all $v$'s neighbors share $v$'s identifier and the BFS-trees rooted at its children are fully constructed. To implement this idea, each node $v$ stores a boolean variable $\texttt{f}_v$ to indicate whether the BFS-tree construction process is completed, and $\texttt{f}_v$ is sent to $v$'s parent in each round. Initially $\texttt{f}_v$ is $false$, and $\texttt{f}_v$ becomes $true$ if: (1) all $v$'s neighbors share identical identifier as $v$; and (2) each child $u$ of $v$ has $\texttt{f}_u=true$ or $v$ has no children (that is, $v$ is a leaf node).

\highlightblue{Lastly, if node $v$ determines that the BFS-tree rooted at itself is fully constructed and it does not have a parent, then it broadcasts a termination signal to its neighbors once and stops the BFS-tree building procedure. The node $v$ will then proceed to the second stage of the algorithm. On the other hand, whenever a node receives a termination signal, it also stops its BFS-tree building procedure, broadcasts this signal to all neighbors once, and then proceed to the second stage of the algorithm.} During algorithm execution, each node $v$ uses a boolean variable $\texttt{build}_v$ to maintain this signal: $\texttt{build}_v$ is initially $true$, and will be set to $false$ when $v$'s BFS-tree building procedure is done. Notice that if there are multiple BFS-trees being constructed simultaneously, after the first one completes, the flooding mechanism of the termination signal may stop the remaining ones from being completed. Nonetheless, such disruption is fine: the existence of multiple BFS-trees implies there are token collisions, and our algorithm can correctly detect this later.

\subparagraph{Detect token collision.} The token-collision detection procedure has two main tasks: compute BFS-tree's size and aggregate tokens. A node only starts this procedure if itself and all its neighbors have terminated the BFS-tree construction procedure.

To determine the size of the BFS-tree rooted at itself, node $v$ first identifies its children. It filters out the neighbors that have the same identifier as itself and have designated node $v$ as their parent. Node $v$ uses $\texttt{chi}_v$ to store this set of children. If node $v$ finds that all of its children in $\texttt{chi}_v$ have already computed the size of their respective BFS-trees, then node $v$ can calculate the size of the BFS-tree rooted at itself.
This is done by summing up the sizes of the BFS-trees rooted at its children and adding one to account for node $v$ itself.
During algorithm execution, each node $v$ uses $\texttt{cnt}_v$ to track the size of the BFS-tree rooted at itself. Initially $\texttt{cnt}_v$ is set to a special symbol $\perp$. Later when $v$ has finished counting, $\texttt{cnt}_v$ becomes an integer.

It remains to aggregate the tokens. In each round, after node $v$ receives all messages (which may include tokens from its children), if $v$ has a parent $u$ and the token list of $v$ is not empty, then $v$ ejects one token from its token list and sends that token to $u$ in the next round. Node $v$ also needs to tell its parent $u$ whether all tokens in the BFS-tree rooted at $v$ has already been transferred to $u$. To this end, in each round, after $v$ has received all messages, if the token list of $v$ is empty and every child of $v$ indicates all tokens have already been transferred to $v$, then $v$ concludes that all tokens in the BFS-tree rooted at itself has already been transferred to its parent. It will inform its parent $u$ about this in the next round. During algorithm execution, each node $v$ uses $\bm{x}^v$ to store its token list and uses $\texttt{ele}_v$ to denote the token that $v$ intends to send to its parent. We note that $\texttt{ele}_v$ is set to $\perp$ when $v$'s token list is empty (that is, $|\bm{x}^v|=0$) and every child $w$ of $v$ indicates all tokens in the subtree rooted at $w$ has been transferred to $v$ (that is, $w$ tells $v$ $\texttt{ele}_w=\perp$); and $\texttt{ele}_v$ is set to $\top$ when $v$'s token list is empty but some child $w$ of $v$ indicates there still are tokens pending to be transferred to $v$ (that is, $w$ tells $v$ $\texttt{ele}_w=\top$).

If node $v$ does not have a parent, it must be the root of some entire BFS-tree and is responsible for deciding the result of token collision.
To this end, once $v$ has obtained the size of the BFS-tree rooted at itself and all its children signal that the tokens have been transferred to $v$, it determines the result of token collision as follows.
In the case that all nodes know the exact value of $n$, if the size of the BFS-tree rooted at $v$ equals $n$ and no token collision is found in the token list of $v$, then $v$ can confirm the non-existence of token collision. Otherwise, a token collision must exist.
In the case that all nodes know the exact value of $k$, if the size of the token list of $v$ equals $k$ and no token collision is found, then $v$ can confirm the non-existence of token collision. Otherwise, a token collision must exist.
Node $v$ uses a boolean variable $\texttt{res}_v$ to store the result. It will broadcast the result to all its neighbors once in the next round and then halt. Upon receiving the result, every node also broadcasts the result to its neighbors once in the next round and then halts.

\section{Analysis of the Deterministic Algorithm}\label{sec:alg-analysis}

In this section, we show the correctness of our algorithm and analyze its running time. Omitted proofs and full version of proof sketches are provided in \Cref{sec-app:det-alg-proof}.

\subsection{Correctness}

We begin with the correctness guarantees: if all nodes know $n$ or $k$, then all nodes return an identical and correct result on whether there are collisions among the $k$ tokens.

To prove the above claim, we first argue the correctness of our BFS-tree construction procedure. Specifically, we intend to show that our algorithm always maintains a directed forest $G'=(V,E')$ where a directed edge $(v,u)\in E'$ if node $v$ assigns node $u$ as its parent. To this end, we introduce the notion of \emph{identifier-induced graph}.

\begin{definition}[\textbf{Identifier-induced Graph}]\label{def:forest}
At the end of any round, define directed graph $G'=(V, E')$ as the \emph{identifier-induced graph} in the following way: $V$ is the node set of the network graph, and a directed edge $(v,u)\in E'$ if $v$ assigns $u$ as its parent.
\end{definition}

To show that the identifier-induced graph is a forest, we begin with the following observation. Recall that each node $v$ uses variable $\rid$ to store its identifier. Intuitively, this lemma holds since node $v$ only updates the value of $\rid_v$ to the value of $\rid_u$ and sets its parent pointer to $u$ when $v$ receives $\rid_u$ from some neighbor $u$ with $\rid_u<\rid_v$.

\begin{lemma}\label{lem:rid-path-increase}
At the end of any round, for any directed path in the identifier-induced graph, the identifiers of the nodes along the directed path are non-increasing.
\end{lemma}

Then, we can show the identifier-induced graph is a directed forest containing one or more rooted trees. To prove the lemma, the key is to show there are no directed cycles in the identifier-induced graph, which can be done by induction on round number.

\begin{lemma}\label{lem:directed-forest}
At the end of any round, the identifier-induced graph is a directed forest in which every weakly connected component is a rooted tree. In particular, in each tree, the unique node with no parent is the root of that tree.\footnote{A weakly connected component of a directed graph is a connected component of the graph when ignoring edge directions.}
\end{lemma}

Next, we show an important property regarding the rooted trees in the identifier-induced graph. Intuitively, it states that within each such tree, nodes may have different identifiers, but for any subtree within the tree, the nodes that have identical identifiers with the root of the subtree are connected and are at the ``top'' of the subtree.

\begin{lemma}\label{lem:subtree}
At the end of any round, for any node $r$, within the subtree rooted at node $r$ in the identifier-induced graph, the subgraph induced by the nodes having identical identifier with node $r$ is also a tree rooted at node $r$.
\end{lemma}

The following key lemma shows that when there are no token collisions, the BFS-tree building procedure constructs a single rooted tree containing all nodes. Though the claim seems straightforward, proving it rigorously turns out to be highly non-trivial.

\begin{lemma}\label{lem:tree-of-dinstinct}
If there are no token collisions, then after all nodes quit the \textsc{BFS-Tree-Building} procedure---that is, after each node $v$ sets $\build_v=false$, the identifier-induced graph contains a single tree rooted at the node having the minimum token as input, and all nodes in that tree have identical identifier.
\end{lemma}

\begin{proof}[Proof sketch]
Throughout the proof, assume there are no token collisions. For each node $v$, let $\vid_v$ denote the minimum token that $v$ received as input. Let $v_{\min}$ denote the unique node having the smallest input token. For any two nodes $u,v\in V$, let $\dist(u,v)$ denote the distance between $u$ and $v$ in the network graph $G$. Define $d=\max_{v\in V}\dist(v,v_{\min})$. For any node $v$, let $d_v$ denote the distance between node $v$ and the nearest node $u$ with $\vid_u$ smaller than $v$. That is, $d_v=\min_{u\in V,\vid_u<\vid_v}\dist(u,v)$. We set $d_{v_{\min}}=+\infty$.

We make the following three claims and prove their correctness via induction on rounds. These claims highlight the key properties our BFS-tree building procedure can enforce.
\begin{enumerate}
	\item \label{item:no-termination-2} No node quits the BFS-tree building procedure within $d$ rounds. Formally, for any $0\leq i\leq d$, each node $v$ has $\build_v=true$ by the end of round $i$.
	
	\item \label{item:each-node-set-rid-2} For any node $v$, any $0\leq i\leq d$, let node $\hat{u}$ be the unique node that has minimum $\vid$ among all nodes $u$ with $\dist(u,v)\leq i$. At the end of round $i$, we have $\rid_v=\vid_{\hat{u}}$. Formally, at the end of round $i$, it holds that $\rid_v=\min_{u\in V, \dist(u,v)\leq i}\vid_u$.
	
	\item \label{item:root-tree-2} For any node $v$, any $0\leq i\leq d$, at the end of round $i$, one of the following cases holds.
	\begin{itemize}
		\item Case I: $d_v>2i$. The nodes with distance at most $i$ to $v$ form a height-$i$ tree rooted at $v$ in the identifier-induced graph. Moreover, for any node $u$ with distance $i$ to $v$, for any node $w$ on the directed path from $u$ to $v$ in the rooted tree, it holds that $\flagnew_w=false$.
		\item Case II: $i<d_v\leq 2i$. There is a tree rooted at $v$ in the identifier-induced graph. Let node $w$ denote the unique node with distance $d_v$ to node $v$ that has the smallest $\vid$. There exists a node $u$ with $\dist(u,v)=d_v-i-1$ and $\dist(u,w)=i+1$ such that $\flagnew$ is $false$ for all nodes along the length-$(d_v-i-1)$ directed path from $u$ to $v$ in the identifier-induced graph.
		\item Case III: $d_v\leq i$. Node $v$ has a parent in the identifier-induced graph.
	\end{itemize}
\end{enumerate}

The claims above easily lead to the lemma. \highlightblue{By \Cref{item:no-termination-2}, no node will quite the \textsc{BFS-Tree-Building} procedure within $d$ rounds.} By \Cref{item:each-node-set-rid-2}, at the end of round $d$, all nodes have identical $\rid$, which is $\vid_{v_{min}}$. By Case I of \Cref{item:root-tree-2}, at the end of round $d$, all nodes form a single tree rooted at $v_{min}$ in the identifier-induced graph. Moreover, no node will ever change parent or $\rid$ later.
\end{proof}

We now proceed to argue the correctness of the token aggregation process, which will lead to the correctness of our entire algorithm.
Recalling \Cref{lem:subtree}, we begin by defining \emph{identifier-induced subtree} to facilitate presentation.

\begin{definition}[\textbf{Identifier-induced Subtree}]\label{def:subtree}
At the end of any round, for any node $r$, within the subtree rooted at $r$ in the identifier-induced graph, call the subtree induced by the nodes that have identical identifier with $r$ as the \emph{identifier-induced subtree rooted at $r$}.
\end{definition}

Our first lemma regarding the correctness of the token aggregation process states that, informally, every root $r$ in the identifier-induced graph correctly computes the size of the identifier-induced subtree rooted at $r$ if it outputs a decision for the token collision problem.

\begin{lemma}\label{lem:cnt-collect}
Assume that in some round $i$, node $v$ runs procedure \textsc{Token-Collision-Detection} and within that procedure updates $\res_v$ for the first time (so that after the update $\res_v\neq\perp$), then by the end of round $i$, the value of $\cnt_v$ equals the size of the identifier-induced subtree rooted at $v$.
\end{lemma}

\begin{proof}[Proof sketch]
Notice that in our algorithm, only a root node in the identifier-induced graph can update its $\res$ within procedure \textsc{Token-Collision-Detection}, so $v$ must be a root node in the identifier-induced graph by the end of round $i$. Let $T_{v,i}$ be the identifier-induced subtree rooted at $v$ by the end of round $i$. We need to show that $\cnt_v$ equals the size of $T_{v,i}$ by the end of round $i$.

To prove the above result, we make the following claim and prove it via an induction on round number: for any node $u$, if $i_u$ is the first round in which $u$ updates $\cnt_u$ to some non-$\perp$ value, then $\cnt_u$ equals the size of $T_{u,i_u}$ by the end of round $i_u$, where $T_{u,i_u}$ is the identifier-induced subtree rooted at $u$ by the end of round $i_u$. Moreover, by the end of any round $i'>i_u$, $\cnt_u$ remains unchanged and $T_{u,i'}$ is identical to $T_{u,i_u}$.

With the above claim, the lemma is easy to obtain. Assume $i_v$ is the first round in which $v$ updates $\cnt_v$ to some non-$\perp$ value, then $\cnt_v$ equals the size of $T_{v,i_v}$ by the end of round $i_v$. Later, at the end of round $i$, when $v$ updates $\res$ to some non-$\perp$ value in procedure \textsc{Token-Collision-Detection}, $\cnt_v$'s value remains unchanged and is $|T_{v,i_v}|=|T_{v,i}|$.
\end{proof}

Our second lemma regarding the correctness of the token aggregation process states that, informally, every root $r$ in the identifier-induced graph correctly collects the tokens within the identifier-induced subtree rooted at $r$ if it outputs a decision for the token collision problem. The high-level strategy for proving this lemma is similar to the proof of \Cref{lem:cnt-collect}, but the details are more involved as the convergecast process is more complicated.

\begin{lemma}\label{lem:token-collect}
Assume that in some round $i$, node $v$ runs procedure \textsc{Token-Collision-Detection} and within that procedure updates $\res_v$ for the first time (so that after the update $\res_v\neq\perp$), then by the end of round $i$, node $v$ collects each token owned by the nodes within the identifier-induced subtree rooted at $v$ exactly once.
\end{lemma}

At this point, we are ready to show the correctness of our algorithm.

\begin{lemma}\label{lem:all-res-same}
After all nodes halt (that is, $\res\neq\perp$), they return the correct result.
\end{lemma}

\begin{proof}
Notice that by algorithm description and \Cref{lem:directed-forest}, only root nodes in the identifier-induced graph can generate $\res\neq\perp$, other nodes can only passively adopt $\res\neq\perp$ from neighbors. So, let $v$ be an arbitrary node that generates $\res\neq\perp$ and assume this happens in round $i_v$, then $v$ must be a root in the identifier-induced graph by the end of round $i_v$. Let $T_{v,i_v}$ denote the identifier-induced subtree rooted at $v$ by the end of round $i_v$.

First consider the case $v$ sets $\res=true$ (that is, there are no token collisions). Then by \Cref{lem:cnt-collect} and \Cref{lem:token-collect}, $v$ has correctly collected the tokens in $T_{v,i_v}$ and correctly counted the size of $T_{v,i_v}$ by the end of round $i_v$, implying that $T_{v,i_v}$ contains all \highlight{input tokens} and there are no collisions among input tokens; that is, the result $v$ generated is correct. Moreover, since $T_{v,i_v}$ is of size $n$, it must be the only tree in the identifier-induced graph. As a result, all nodes other than $v$ will only passively adopt the result generated by $v$, implying that all nodes return identical result.

Next, consider the case $v$ sets $\res=false$ (that is, there are token collisions). Then, again, by \Cref{lem:cnt-collect} and \Cref{lem:token-collect}, $v$ has correctly collected the tokens in $T_{v,i_v}$ and correctly counted the size of $T_{v,i_v}$ by the end of round $i_v$. Since $v$ sets $\res$ to $false$, there are three possible reasons:
\begin{itemize}
	\item All nodes know $n$ but $|T_{v,i_v}|\neq n$. Recall \Cref{lem:tree-of-dinstinct}, which states that if there are no token collisions, then there is only one tree in the identifier-induced graph that contains all nodes and all tokens. Hence, if $|T_{v,i_v}|\neq n$, then there are indeed token collisions. Furthermore, for any other root node $u$ in the identifier-induced graph that also generates an $\res\neq\perp$ by the end of some round $i_u$, node $u$ must have also found $|T_{u,i_u}|\neq n$ and set $\res=false$. Therefore, in this case, all nodes output the correct result.
	\item All nodes know $k$ but the number of tokens $v$ has collected is not $k$. By a similar argument as in the first case, we can conclude that all nodes output the correct result.
	\item Node $v$ finds collisions among the tokens it has collected. In this case, $v$'s decision to generate $\res=false$ is obviously correct. Moreover, for any other root node $u$ in the identifier-induced graph that also generates an $\res\neq\perp$, that $\res$ must be $false$, as the identifier-induced subtree rooted at $u$ will not contain all $n$ nodes or all $k$ tokens.
\end{itemize}
This completes the proof of the lemma.
\end{proof}

\subsection{Complexity}

We now proceed to analyze the time complexity of the algorithm. The first lemma states that any identifier-induced subtree rooted at some node that has a global minimum token as input has limited height---particularly, $O(D)$. Moreover, each such node is a root in identifier-induced graph.

\begin{lemma}\label{lem:comp-min-subtree-D}
Let $v$ be a node having a minimum token as input. Then at the end of any round, $v$ is a root in the identifier-induced graph, and the identifier-induced subtree rooted at $v$ has height $O(D)$.
\end{lemma}

With \Cref{lem:comp-min-subtree-D}, we argue that all nodes finish \textsc{BFS-Tree-Building} in $O(D)$ rounds.

\begin{lemma}\label{lem:comp-all-quit-BFS}
After $O(D) $ rounds, every node $v$ quits \textsc{BFS-Tree-Building} (that is, $\build_v=false$).
\end{lemma}

\begin{proof}
To prove the lemma, we only need to show that some node will quit \textsc{BFS-Tree-Building} within $O(D)$ rounds. This is because the flooding mechanism of a $false$-valued $\build$ variable ensures, once a node $v$ sets $\build_v=false$, all other nodes will set $\build$ to $false$ within (at most) another $D$ rounds.

If some node quits \textsc{BFS-Tree-Building} within $D$ rounds then we are done, so assume that this is not true. Then, by the end of round $D$, global minimum token's value is known by every node. Particularly, by the end of round $D$, each node has an $\rid$ with a value equals to some global minimum token. In other words, by the end of round $D$, each node is in some identifier-induced subtree rooted at some node that has a global minimum token as input. Moreover, no node will change its $\rid$ or parent ever since. By \Cref{lem:comp-min-subtree-D}, any tree rooted at some node that has a global minimum token as input has $O(D)$ height.

Now, by our algorithm, starting from round $D+1$, nodes within any such tree will start setting $\flagnew$ to $true$ from leaves to root. Since the height of any such tree is $O(D)$, after $O(D)$ rounds, either some node already sets $\build$ to $false$ and quits \textsc{BFS-Tree-Building}, or some root of such tree sets $\build$ to $false$ and quits \textsc{BFS-Tree-Building}. In both cases, some node quits \textsc{BFS-Tree-Building} within $O(D)$ rounds since the start of execution.
\end{proof}

The next lemma states the time complexity of our algorithm.

\begin{lemma}\label{lem:comp-all-quit-decide}
After $O(D+k)$ rounds, every node $v$ halts (that is, $v$ returns $\res\neq\perp$).
\end{lemma}

\begin{proof}[Proof sketch]
Recall that our algorithm guarantees that if one node generates an $\res\neq\perp$ and then halts, then this $\res$ is broadcast to all other nodes. Hence, all other nodes will halt within another $D$ rounds. As a result, to prove the lemma, we show that some node will halt within $O(D+k)$ rounds. To this end, we show that after all nodes quit \textsc{BFS-Tree-Building} which happens within $O(D)$ rounds (by \Cref{lem:comp-all-quit-BFS}), there exists a tree in the identifier-induced graph of height $O(D)$ (by \Cref{lem:comp-min-subtree-D}), and the convergecast process inside this tree take $O(D+k)$ rounds.
\end{proof}

\subsection{Proof of the main theorem}

We now prove \Cref{thm:upper-det-part1}.
When $L=\Theta(\log{n})$---meaning that each message can fit at most a constant number of tokens, by \Cref{lem:all-res-same} and \Cref{lem:comp-all-quit-decide}, the theorem is immediate.

When $L=o(\log{n})$, to prove the theorem, we make a small modification to our algorithm: in the convergecast process, whenever a node forwards tokens to its parent, it packs as many tokens in a message as possible (particularly, $\Theta((\log{n})/L)$ tokens in a message). Intuitively, this means that our algorithm is convergecasting $\Theta(kL/\log{n})$ ``packed tokens'' each of size $\Theta(\log{n})$, and each of these ``packed tokens'' contains $\Theta((\log{n})/L)$ real tokens. Hence, the total runtime of our algorithm is still $O(D+kL/\log{n})$ rounds.

The above argument is valid if, for every node that has some token(s) as input, that node receives at least $\Theta((\log{n})/L)$ tokens. If some node only receives $o((\log{n})/L)$ tokens as input (e.g., only one token), then a more careful analysis is required.
Specifically, assume that there are $x$ nodes that each receives $o((\log{n})/L)$ tokens as input, call these nodes $V_x$, and the nodes in $V_x$ in total have $k_x$ tokens. So, there are $n-x$ nodes that each receives at least $\Theta((\log{n})/L)$ tokens as input, call these nodes $V_{\overline{x}}$, and the nodes in $V_{\overline{x}}$ in total have $k-k_x$ tokens. Imagine a process in which we first aggregate the tokens owned by $V_{\overline{x}}$, and then aggregate the tokens owned by $V_x$. By the above analysis, aggregating the tokens owned by $V_{\overline{x}}$ takes $O(D+(k-k_x)L/\log{n})$ rounds. On the other hand, for each token owned by some node in $V_x$, within $O(D)$ rounds, it either reaches the root, or arrives at a node that has at least $\Theta((\log{n})/L)$ tokens pending to be sent. Effectively, this means that starting from the round we process the tokens owned by $V_x$, in $O(D)$ rounds, we again arrive at a scenario in which each node that has pending tokens to send has at least $\Theta((\log{n})/L)$ tokens in its token list. As a result, these $k_x$ tokens owned by $V_x$ will all reach the root within $O(D+k_xL/\log{n})$ rounds. Note that our modified algorithm cannot be slower than the imagined process, so the runtime of our modified algorithm when $L=o(\log{n})$ is $O(D+kL/\log{n})$.

\section{Generalizing the Deterministic Algorithm when Tokens are Large}\label{sec:alg-extension}

When tokens are large, $L=\omega(\log{n})$ in particular, the time complexity of the BFS-tree building process and the token aggregation process are both affected. As mentioned in \Cref{sec:intro}, we can apply the simple strategy of using $L/\log{n}$ rounds to simulate one round of our algorithm (as a token can be transferred in $L/\log{n}$ rounds), but the resulting algorithm would be too slow. Instead, in this section, we introduce and analyze a variant of our algorithm that costs only \highlight{$O(D\cdot\max\{\frac{\log(L/\log n)}{\log n},1\}+k\cdot\frac{L}{\log n})$} rounds when $L=\omega(\log{n})$.

The high level framework of this variant is the same as the algorithm introduced in \Cref{sec:alg-det}: first build BFS-tree(s) and then detect token collisions within the tree(s). In this section, we focus on introducing the process of building BFS-tree(s) as the latter component is almost identical with the original algorithm.
(Complete pseudocode of this variant is provided in \Cref{sec-app:pseudocode}.) \highlightblue{For the ease of presentation, we use $B=\Theta(\log{n})$ to denote the bandwidth of CONGEST networks throughout this section.}

\subsection{Algorithm description}

We first explain the key idea that allows this variant to be faster than the simulation strategy. Recall that in the original BFS-tree building process, each node $v$ needs to record the minimum token it has seen in $\rid_v$, and this is done by exchanging tokens in their \emph{entirety} with neighbors. However, a key observation is, the relative order of two binary strings can be determined by a \emph{prefix} of the strings that includes the most significant bit where they differ.
As a result, we can employ the strategy that identifiers are sent successively starting from the most significant bit. Whenever a node $v$ finds a prefix from some neighbor $u$ is strictly smaller than the prefix of its current identifier, $v$ updates its identifier to match the prefix and designates $u$ as its parent. Moreover, when $v$ sends its updated identifier, it does not need to restart from the first bit; instead, $v$ starts from the bit where the updated identifier differs from the previous identifier.
Effectively, we obtain an efficient ``\emph{pipeline}'' approach on identifier broadcasting that can speed up the BFS-tree building process.

\subparagraph{Build BFS-tree(s).} We now detail how to implement the above idea. Similar to the original algorithm, each node $v$ attempts to construct a BFS-tree rooted at itself by broadcasting its identifier $\rid_v$. Due to bandwidth limitation, each identifier is divided into multiple \emph{pieces} so that one piece can fit into one message. \highlightblue{Denote these pieces as $\rid_v[1],\cdots,\rid_v[\lceil L/B\rceil]$, where $\rid_v[1]$ contains the $B$ most significant bits while $\rid_v[\lceil L/B\rceil]$ contains the $B$ least significant bits.}  The BFS-tree building procedure contains multiple \emph{iterations}, each of which contains \highlightblue{$\Theta(\frac{\log{(L/B)}}{B})$} rounds. In each iteration, \highlightblue{$\Theta(\frac{\log{(L/B)}}{B})$} identifier pieces are sent, \highlightblue{along with the position of the first sent piece in $\rid_v$}---we use $\sent_v$ to denote this position. \highlightblue{(Notice that sending $\sent_v$ may require $\frac{\log{(L/B)}}{B}$ rounds when $L$ is large, this is why each iteration may contain multiple rounds.)}
Each node $v$ locally maintains an identifier prefix for each neighbor based on received pieces. Whenever $v$ finds a prefix of some neighbor $u$ is strictly smaller than the prefix of its current identifier, $v$ updates its identifier to match the prefix and designates $u$ as its parent. At this point, $v$ should send the updated identifier to neighbors. Particularly, $v$ starts with the first piece where the updated identifier differs from $v$'s previous identifier.
This implies $v$ may send non-successive piece position, in which case each neighbor of $v$ should abandon the old prefix of $v$ and record the new one.

Node $v$ waits until all neighbors and itself have sent complete identifiers. Then, if $v$ finds that all neighbors share the same identifier as itself, it attempts to ascertain whether the BFS-tree rooted at itself is fully constructed. Similar to the original algorithm, each node $v$ uses a boolean variable $\texttt{f}_v$ to indicate whether BFS-tree construction is completed. Initially $\texttt{f}_v$ is $false$, and $\texttt{f}_v$ becomes $true$ if: (1) $v$ and all its neighbors have sent complete identifiers; (2) $v$ and all its neighbors have identical identifier; and (3) each child $u$ of $v$ has $\texttt{f}_u=true$ or $v$ has no children.

Lastly, if node $v$ determines that the BFS-tree rooted at itself is fully constructed and it does not have a parent, then it terminates the BFS-tree building procedure and broadcasts a termination signal to all neighbors once. \highlightblue{The node will then proceed to the second stage of the algorithm.} Any node receiving such a signal will also forward it to neighbors once, stop the BFS-tree building procedure, and proceed to the second stage of the algorithm.

\subsection{Analysis}

The analysis for the above generalized algorithm is similar to the analysis for the original algorithm.
Most claims and lemmas can carry over with little or no modifications, so are the proofs for these claims and lemmas. \highlightblue{Others, however, require non-trivial extension or adjustments.}
To avoid redundancy, we only state these claims and lemmas here and provide proofs that require noticeable extension or adjustments in \Cref{sec-app:det-alg-proof-gen}.

\subparagraph{Correctness.} The definition for identifier-induced graph remains unchanged in the generalized setting, except that such graph is defined at the end of each iteration.

\begin{definition}[Analogue of \Cref{def:forest}]\label{def:forest-gen}
At the end of any iteration, define directed graph $G'=(V, E')$ as the \emph{identifier-induced graph} in the following way: $V$ is the node set of the network graph, and a directed edge $(v,u)\in E'$ if $v$ assigns $u$ as its parent.
\end{definition}

Following lemma is an analogue of \Cref{lem:rid-path-increase}, its proof is almost identical to that of \Cref{lem:rid-path-increase}, with small adjustments to account for the fact that identifiers are sent in pieces.

\begin{lemma}[Analogue of \Cref{lem:rid-path-increase}]\label{lem:rid-path-increase-gen}
At the end of any iteration, for any directed path in the identifier-induced graph, the identifiers of the nodes along the directed path are non-increasing.
\end{lemma}

With \Cref{lem:rid-path-increase-gen}, analogues of \Cref{lem:directed-forest} and \Cref{lem:subtree} hold automatically.

\begin{lemma}[Analogue of \Cref{lem:directed-forest}]\label{lem:directed-forest-gen}
At the end of any iteration, the identifier-induced graph is a directed forest in which every weakly connected component is a rooted tree. In particular, in each tree, the unique node with no parent is the root of that tree.
\end{lemma}

\begin{lemma} [Analogue of \Cref{lem:subtree}]\label{lem:subtree-gen}
At the end of any iteration, for any node $r$ that has sent its complete identifier to neighbors (that is, \highlightblue{$\sent_r=\ceillb$}), within the subtree rooted at node $r$ in the identifier-induced graph, the subgraph induced by the nodes having identical identifier with node $r$ is also a tree rooted at node $r$.
\end{lemma}

\Cref{lem:tree-of-dinstinct} is critical for the original algorithm, which states that a single BFS tree containing all nodes will be built when there are no token collisions. In the generalized setting, this claim still holds, but the proof needs to be extended in a non-trivial fashion to deal with the complication introduced by the pipeline approach for sending identifiers.

\begin{lemma} [Analogue of \Cref{lem:tree-of-dinstinct}]\label{lem:tree-of-dinstinct-gen}
If there are no token collisions, then after all nodes quit the \textsc{BFS-Tree-Building} procedure, the identifier-induced graph contains a single tree rooted at the node having the minimum token as input, and all nodes in that tree have identical identifier.
\end{lemma}

Much like the case of \Cref{def:forest-gen}, the definition for identifier-induced subtree remains largely unchanged in the generalized setting.

\begin{definition}[Analogue of \Cref{def:subtree}]\label{def:subtree-gen}
At the end of any iteration, for any node $r$ that has sent its complete identifier to neighbors (that is, \highlightblue{$\sent_r=\ceillb$}), within the subtree rooted at $r$ in the identifier-induced graph, call the subtree induced by the nodes having identical identifier with $r$ as the \emph{identifier-induced subtree rooted at $r$}.
\end{definition}

\Cref{lem:cnt-collect} and \Cref{lem:token-collect} (and their proofs) still hold in the generalized setting, as we utilize the mechanism in the original algorithm for counting tree size and aggregating tokens.

\begin{lemma} [Analogue of \Cref{lem:cnt-collect}]\label{lem:cnt-collect-gen}
Assume that in some iteration $i$, node $v$ runs procedure \textsc{Token-Collision-Detection} and within that procedure updates $\res_v$ for the first time (so that after the update $\res_v\neq\perp$), then by the end of iteration $i$, the value of $\cnt_v$ equals the size of the identifier-induced subtree rooted at $v$.
\end{lemma}

\begin{lemma}[Analogue of \Cref{lem:token-collect}]\label{lem:token-collect-gen}
Assume that in some iteration $i$, node $v$ runs procedure \textsc{Token-Collision-Detection} and within that procedure updates $\res_v$ for the first time (so that after the update $\res_v\neq\perp$), then by the end of iteration $i$, node $v$ collects each token owned by the nodes within the identifier-induced subtree rooted at $v$ exactly once.
\end{lemma}

We conclude this part with the following lemma which shows the correctness of our generalized algorithm, its proof is essentially identical to that of \Cref{lem:all-res-same}.

\begin{lemma}[Analogue of \Cref{lem:all-res-same}]\label{lem:all-res-same-gen}
After all nodes halt (that is, $\res\neq\perp$), they return identical and correct result.
\end{lemma}

\subparagraph{Complexity.} We now analyze the round complexity of the generalized algorithm, focusing on the BFS-tree construction process. Firstly, an analogue of \Cref{lem:comp-min-subtree-D} can be established.

\begin{lemma}[Analogue of \Cref{lem:comp-min-subtree-D}]\label{lem:comp-min-subtree-D-gen}
Let $v$ be a node having a minimum token as input. At the end of any iteration, if $v$ has sent its complete identifier to neighbors (that is, \highlightblue{$\sent_v=\ceillb$}), then $v$ is a root in the identifier-induced graph, and the identifier-induced subtree rooted at $v$ has height $O(D)$.
\end{lemma}

The next lemma states the time consumption of the BFS-tree construction process, it highlights the advantage of using the pipelining approach over the simulation approach.

\begin{lemma}[Analogue of \Cref{lem:comp-all-quit-BFS}]\label{lem:comp-all-quit-BFS-gen}
After \highlightblue{$O(D + \frac{L}{\log(L/\log n)})$} iterations, every node $v$ quits \textsc{BFS-Tree-Building} (that is, $\build_v=false$).
\end{lemma}

The last lemma shows the total time complexity of the generalized algorithm.

\begin{lemma} [Analogue of \Cref{lem:comp-all-quit-decide}]\label{lem:comp-all-quit-decide-gen}
After \highlightblue{$O(D\cdot\frac{\log(L/\log n)}{\log n} + k\cdot\frac{L}{\log n})$} rounds, every node $v$ halts (that is, $v$ returns $\res\neq\perp$).
\end{lemma}

\subparagraph{Proof of the main theorem.} Combine \Cref{lem:all-res-same-gen} and \Cref{lem:comp-all-quit-decide-gen}, \Cref{thm:upper-det-part2} is immediate.

\section{Impossibility Result and Lower Bound for Deterministic Algorithms}\label{sec:lower-bound}

\subparagraph{Impossibility result.} Recall \Cref{thm:impossible} which states that if each node has no knowledge about the network graph except being able to count and communicate over adjacent links, and if each node also has no knowledge regarding the tokens except the ones being given as input, then the token collision problem has no deterministic solution.

To obtain the above impossibility, the key intuition is: to solve the problem, nodes need to exchange their input tokens in some manner; but in the anonymous setting with no global knowledge regarding network graph or input tokens, whenever a node receives a token from some neighbor that collide with its own input, the node cannot reliably determine whether this token originates from itself or some other node, yet the correctness of any algorithm depends on being able to distinguish these two scenarios.

We now provide a complete proof. Note that our impossibility result is strong in that we can construct counterexamples for any network size $n\geq 3$.

\begin{proof}[Proof of \Cref{thm:impossible}]
Assume that there is an algorithm $\mathcal{A}$ that solves the token collision problem in the considered setting.
For any $n\geq 3$, we consider two problem instances. The first instance---henceforth called $C_n$---is a ring consisting of $n$ nodes, denoted as $v_1,v_2,\cdots,v_n$. Each node in the network obtains one token as input. Particularly, for any $i\in[n]$, node $v_i$ has a token with value $i$. The second instance---henceforth called $C_{2n}$---is a ring consisting of $2n$ nodes. To construct $C_{2n}$, we first build two paths. The first path contains $n$ nodes, denoted as $v'_1,v'_2,\cdots,v'_n$; the second path also contains $n$ nodes, denoted as $u_1,u_2,\cdots,u_n$. Then, we connect $v'_n$ with $u_1$, and connect $u_n$ with $v'_1$. At this point, we have a ring. Each node in $C_{2n}$ obtains one token as input. Particularly, for any $i\in[n]$, node $v'_i$ and node $u_i$ each has a token with value $i$.
\Cref{fig:impossible-result-1} shows an example of the two instances when $n=3$.

\begin{figure}[ht!]
\centering
\includegraphics[scale=0.25]{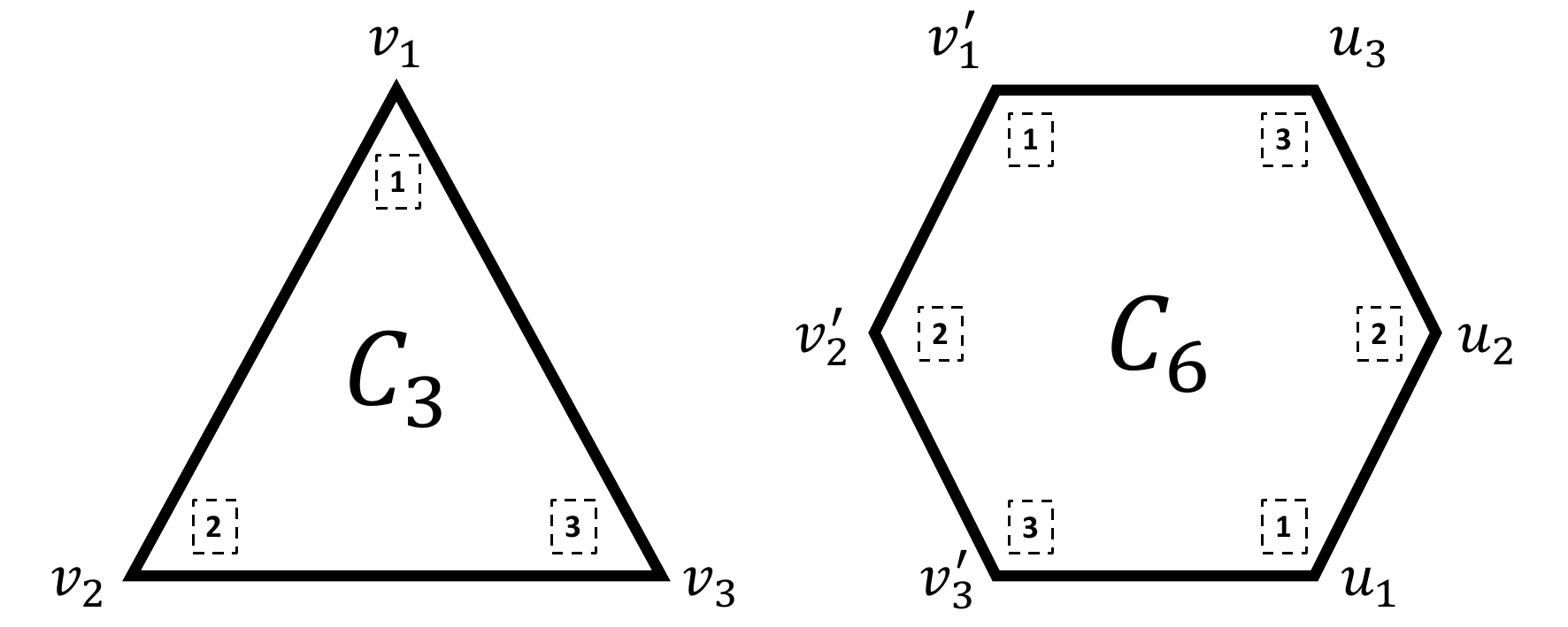}
\vspace{-2ex}
\caption{Example instances for the impossibility proof ($n=3$), nodes' input are in dashed box.}\label{fig:impossible-result-1}
\end{figure}

Clearly, token collisions exist in $C_{2n}$ but not in $C_{n}$, yet we will prove that $\mathcal{A}$ outputs identical results in both instances, resulting in a contradiction. Specifically, call the execution of $\mathcal{A}$ on $C_n$ as $\alpha$ and the execution of $\mathcal{A}$ on $C_{2n}$ as $\beta$, we will prove by induction that by the end of every round, for any $i\in[n]$, the internal states of nodes $v_i$, $v'_i$ and $u_i$ are identical.

The base case, which is immediately after initialization (i.e., round 0), trivially holds.

Assume that the claim holds for all rounds up to the end of round $r\geq 0$, now consider round $r+1$. Fix an arbitrary $i\in[n]$, by the induction hypothesis, nodes $v_{i-1}$ and $v'_{i-1}$ have identical states by the end of round $r$. Hence, in round $r+1$, the message (if any) $v_{i-1}$ sends to $v_i$ and the message (if any) $v'_{i-1}$ sends to $v'_i$ will be identical. Similarly, in round $r+1$, the message (if any) $v_{i+1}$ sends to $v_i$ and the message (if any) $v'_{i+1}$ sends to $v'_i$ will be identical. Also, notice that by the end of round $r$, by the induction hypothesis, $v_i$ and $v'_i$ have identical states. Hence, during round $r+1$, the local views of $v_i$ and $v'_i$ are identical. In other words, for any $\hat{v}\in\{v_i,v'_i\}$, node $\hat{v}$ cannot distinguish whether it is in $\alpha$ or $\beta$. Therefore, by the end of round $r+1$, nodes $v_i$ and $v'_i$ have identical states. By a similar argument, we can show that by the end of round $r+1$, nodes $v_i$ and $u_i$ also have identical states. This completes the proof of the inductive step, hence proving the claim.

Since $\mathcal{A}$ solves the token collision problem, $\alpha$ and $\beta$ both terminate. Moreover, due to the above claim, nodes in $\alpha$ and $\beta$ output identical results, resulting in a contradiction.
\end{proof}

\subparagraph{Deterministic lower bound.} We reduce the set-disjointness problem to the token collision problem and obtain the following theorem. \Cref{thm:lower-det} is an immediate corollary of it (by setting mincut$(G)=1$).

\begin{theorem}\label{thm:lower-det-raw}
Recall the parameters $n,k,L$ introduced in the definition of the token collision problem (that is, \Cref{def:problem}). Consider a size-$n$ \highlightblue{CONGEST} network $G=(V,E)$ with diameter $D$.
Assuming $2^L\ge k$, any deterministic algorithm that solves the token collision problem takes \highlight{$\Omega(D+\frac{k(L-\log{k}+1)}{{\rm mincut}(G)\cdot \log n})$} rounds.
Here, ${\rm mincut(G)}$ denotes the mincut of $G$.
That is, ${\rm mincut}(G)=\min_{U\subset V}|\{(u,v)\in E~|~u\in U,v\in V\setminus U\}|$.
\end{theorem}

\begin{proof}
Let $(U,V\setminus U)$ be a partition of $V$ that attains ${\rm mincut}(G)$. Let $u\in U$ and $v\in V\setminus U$ be a pair of farthest nodes between $U$ and $V\setminus U$. Assume that $u$ and $v$ are assigned token sets $S$ and $T$ respectively, each containing $k/2$ tokens.

Notice that $\Omega(D)$ is a lower bound for the token collision problem, as the distance between $u$ and $v$ is $\Theta(D)$, and they need to communicate with each other to solve the problem.

On the other hand, recall the two-party communication model and the set-disjointness problem introduced in \Cref{sec:preliminary}. Since $2^L\ge k$, by setting $p=2^L$ and $q=k/2$, it holds $q\le p/2$. \highlightblue{Recall the bandwidth of the network is $B=\Theta(\log n)$.} We claim, if there exists an $r$-round algorithm that deterministically solves token collision in the CONGEST model, then Alice and Bob can compute ${\rm DISJ}^p_q(S,T)$ by communicating at most $2rB\cdot{\rm mincut}(G)$ bits. Specifically, they can run the $r$-round algorithm by having Alice and Bob simulate nodes in $U$ and $V\setminus U$ respectively. Communication between Alice and Bob is necessary only when messages (each of which is at most $B$ bits) are exchanged between nodes in $U$ and $V\setminus U$ in the simulation. Apply \Cref{fact:disj}, we have $2r\cdot{\rm mincut}(G)\cdot B=\Omega(\log\tbinom pq)$, implying \highlight{$r=\Omega(\frac{k(L-\log k+1)}{{\rm mincut}(G)\cdot B})=\Omega(\frac{k(L-\log k+1)}{{\rm mincut}(G)\cdot \log n})$}.
\end{proof}

\section{The Randomized Scenario}\label{sec:rnd-scenario}

For the sake of completeness, in this section, we briefly discuss the round complexity of the token collision problem when randomization is allowed.

\subparagraph{Randomized upper bound.} We first describe a randomized algorithm that solves token collision with probability at least $1-1/k$ within \highlightblue{$O(D\cdot\frac{\log((\log{k})/\log n)}{\log n} + k\cdot\frac{\log{k}}{\log n} + \frac{L}{\log n})$} rounds, hence proving the upper bound part of \Cref{thm:rnd-upper-lower}.

To begin with, we elect a leader among the nodes that have at least one token as input. Notice that there are at most $k$ such nodes. Hence, by letting each such node $v$ uniformly and independently sample $\vid_v\in\{0,1\}^{c\log k}$ for some sufficiently large constant $c$, there is a unique node $\hat{v}$ that obtains the global minimum $\vid$ with probability at least $1-1/k^{c-2}$. If we let each node continuously broadcast the minimum $\vid$ that it ever received, a size-$n$ BFS tree rooted at $\hat{v}$ would be constructed with probability at least $1-1/k^{c-2}$. Moreover, by using the pipelining approach we introduced in \Cref{sec:alg-extension}, this process takes at most \highlightblue{$O(D\cdot\frac{\log((\log{k})/\log n)}{\log n}+\frac{\log{k}}{\log n})$} rounds (see \Cref{lem:comp-all-quit-BFS-gen}).

Once a size-$n$ BFS tree is built, the root---which is also the leader---will collect all tokens to determine whether collisions exist. Notice that with randomization, we do not have to transfer each token in its entirety. In particular, we can leverage the following fact on the collision probability of random hash function to reduce the length of each token.

\begin{fact}[\cite{brody14}]
For any set $S\subseteq[2^L]$ of size $|S|=k$ and any $\beta\ge0$, there exists a random hash function $h:[2^L]\to[q]$ with $q=O(k^{2+\beta})$ such that, with probability at least $1-1/k^\beta$, it holds that $h(x)\ne h(y)$ for all $x,y\in S$ with $x\ne y$.
Moreover, $h$ can be constructed using $O(L)$ random bits.
\end{fact}

Therefore, after BFS-tree construction, the leader can generate $O(L)$ random bits for constructing the random hash function, and broadcast these bits to all nodes in $O(D+\frac{L}{\log n})$ rounds. Then, each node uses the random hash function to reduce the length of its tokens to $O(\log{k})$ bits. Finally, the $k$ tokens each of length $O(\log{k})$ is aggregated to the root in $O(D+k\cdot\frac{\log{k}}{\log n})$ rounds.

Clearly, the total runtime of the algorithm is
$$O\left(D\cdot\frac{\log((\log{k})/\log n)}{\log n} + k\cdot\frac{\log{k}}{\log n} + \frac{L}{\log n}\right),$$
and it succeeds with probability at least $1-1/k$.

\subparagraph{Randomized lower bound.} The lower bound part of \Cref{thm:rnd-upper-lower} can be obtained in the same manner as the proof of \Cref{thm:lower-det-raw} by using the randomized lower bound of set-disjointness mentioned in \Cref{fact:disj}. We omit its proof to avoid redundancy.


\bibliography{disc24-draft-v4}


\appendix

\section*{Appendix}

\section{Pseudocode of the Deterministic Algorithms}\label{sec-app:pseudocode}

The complete pseudocode of the algorithm in \Cref{sec:alg-det} is given in \Cref{fig:alg-main}. Here are the explanations of some key variables that are used in the pseudocode. For any node $v$,
\begin{itemize}
	\item $\build_v$: a boolean variable indicating whether BFS-tree building is ongoing for $v$.
	\item $\rid_v$: the identifier of $v$, intuitively it stores the root of the BFS-tree that $v$ belongs to.
	\item $\parent_v$: the label of the edge connecting to the parent of $v$.
	\item $\child_v$: the set of edge labels representing the children of $v$.
	\item $\flagnew_v$: a boolean variable indicating whether the BFS-tree rooted at $v$ is fully constructed.
	\item $\cnt_v$: the size of the BFS-tree rooted at $v$.
	\item $\ele_v$: if $\ele_v\notin\{\perp,\top\}$, then it is the token that $v$ intends to send to its parent in the next round; if $\ele_v=\top$, it indicates that there may exist a token in the BFS-tree rooted at $v$ that has not been transferred to $v$'s parent; if $\ele_v=\perp$, it indicates that all tokens in the BFS-tree rooted at $v$ have already been transferred to $v$'s parent.
	\item $\res_v$: the result of the token collision problem, that is, the algorithm's output at node $v$.
\end{itemize}

\begin{figure}[ht!]
\hrule
\vspace{1ex}
{Main algorithm executed at each node $v$.}
\vspace{.5ex}
\hrule
\begin{small}
\begin{algorithmic}[1]
\State $\build_v \gets true$, $\bm{x}^v\gets v\text{'s input tokens}$, $\rid_v \gets \min\{\bm{x}^v\} $, $\parent_v \gets \perp$, $\child_v \gets \perp$, $\flagnew_v\gets false$.
\State $\cnt_v\gets \perp$, $\ele_v\gets\top$, $\res_v\gets \perp$.  \Comment{End of initialization.}
\For{(each round)}
	\For{(each incident edge with label $i\in[\Delta_v]$)}
		\State $\ischild_i\gets\mathbb{I}[\parent_v==i]$. \Comment{$\ischild_i$ indicates whether edge $i$ connects to the parent of $v$.}
		\State Send $\langle \res_v, \build_v, \rid_v, \ischild_i, \flagnew_v, \cnt_v, \ele_v\rangle$ through edge $i$.
	\EndFor
	\If{($\res_v\neq \perp$)}
		Return $\res_v$ as final result. \Comment{Termination.}
	\EndIf
	\State For {\scriptsize $i\in [\Delta_v]$}, let {\scriptsize $m_i=\langle \res_i, \build_i, \rid_i, \ischild_i, \flagnew_i, \cnt_i, \ele_i \rangle$} be the message received via edge $i$.
	\For{(each edge $i\in [\Delta_v]$)}
		\If{($\res_i\neq \perp$)}
			$\res_v\gets \res_i$.
		\EndIf
		\State $\build_v \gets \build_v \wedge \build_i$.
	\EndFor
\If{($\build_v==true$)}
	Execute Procedure \textsc{BFS-Tree-Building}.
\Else
	\xspace Execute Procedure \textsc{Token-Collision-Detection}.
\EndIf
\EndFor
\end{algorithmic}
\end{small}
\hrule
\vspace{3ex}
\hrule
\vspace{1ex}
{Procedure \textsc{BFS-Tree-Building} executed at node $v$.}
\vspace{1ex}
\hrule
\begin{small}
\begin{algorithmic}[1]
\State $ID_v\gets \{\rid_i \mid  i\in [\Delta_v]\}$.
\If{($\min\{ID_v\}<\rid_v$)}
	\State Let $j\in [\Delta_v]$ be one edge label satisfying $\rid_j==\min\{ID_v\}$.
	\State $\rid_v\gets \rid_j$, $\parent_v\gets j$, $\flagnew_v\gets false$. \Comment{Notice that $\flagnew_v$ is reset to $false$.}\label{line:update-rid}
\ElsIf{($\max\{ID_v\}==\rid_v$)}
	\State $\child_v\gets \{i\mid i\in [\Delta_v] \text{ and }\ischild_i==true\}$.
	\If{(($\forall i\in \child_v$, $\flagnew_i==true$) \textbf{or} $\child_v==\emptyset$)}
		$\flagnew_v\gets true$.
	\EndIf
\EndIf
\If{($\parent_v == \perp$ \textbf{and} $\flagnew_v == true$)}
	$\build_v\gets false$.
\EndIf
\end{algorithmic}
\end{small}
\hrule
\vspace{3ex}
\hrule
\vspace{1ex}
{Procedure \textsc{Token-Collision-Detection} executed at node $v$.}
\vspace{1ex}
\hrule
\begin{small}
\begin{algorithmic}[1]
\State \highlightblue{$\child_v\gets \{i\mid i\in [\Delta_v] \text{ and } \build_i==false \text{ and } \ischild_i== true \text{ and } \rid_i==\rid_v\}$.}
\State \highlightblue{Append $\{\ele_i\mid i\in \child_v\text{ and } \ele_i\in\{0,1\}^L\}$ to $\bm{x}^v$.}
\If{($\forall i\in [\Delta_v]$, $\build_i==false$)}
	\If{(($\forall i\in \child_v$, $\cnt_i\neq \perp$) \textbf{or} $\child_v==\emptyset$)}
		$\cnt_v\gets 1+\sum_{i\in \child_v} \cnt_i$.
	\EndIf
	\If{($\parent_v\neq \perp$)}
		\If{($|\bm{x}^v|>0$)}
			Eject one token from $\bm{x}^v$ and let that token be $\ele_v$.
		\ElsIf{(($\forall i\in \child_v$, $\ele_i==\perp$) \textbf{or} $\child_v==\emptyset$)}
			$\ele_v\gets \perp$.
		\Else
			\xspace $\ele_v\gets \top$.
		\EndIf
	\ElsIf{($\cnt_v\neq \perp$ \textbf{and} (($\forall i\in \child_v$, $\ele_i== \perp$) \textbf{or} $\child_v==\emptyset$))}
		\If{(know value of $n$ \textbf{and} $\cnt_v== n$ \textbf{and} no token collision in $\bm{x}^v$)}
			$\res_v\gets true$.
		\ElsIf{(know value of $k$ \textbf{and} $|\bm{x}^v|==k$ \textbf{and} no token collision in $\bm{x}^v$)}
			$\res_v\gets true$.
		\Else
			\xspace $\res_v\gets false$.
		\EndIf
	\EndIf
\EndIf
\end{algorithmic}
\end{small}
\hrule
\vspace{2ex}
\caption{Pseudocode of the deterministic token collision algorithm.}\label{fig:alg-main}
\vspace{-4ex}
\end{figure}

\smallskip\noindent The pseudocode of the algorithm introduced in \Cref{sec:alg-extension}, which deals with the case that tokens are large, are given in \Cref{fig:alg-main-alt} and \Cref{fig:alg-procedures-alt}.

\begin{figure}[t]
\hrule
\vspace{1ex}
{Main algorithm executed at each node $v$ for large tokens.}
\vspace{.5ex}
\hrule
\begin{small}
\begin{algorithmic}[1]
\State $\build_v \gets true$, $\bm{x}^v\gets v\text{'s input tokens}$, $\rid_v \gets \min\{\bm{x}^v\}$, $\parent_v \gets \perp$, $\child_v \gets \perp$, $\flagnew_v\gets false$.
\State $\cnt_v\gets \perp$, $\ele_v\gets \top$, $\res_v\gets \perp$, $\sent_v\gets 0$.
\State Initialize four vectors $\{\rid_i\}_{i\in [ \Delta_v]}$, $\{\sent_i\}_{i\in [\Delta_v]}$, $\{\ele_i\}_{i\in \{0,\cdots, \Delta_v\}}$, and $\{\sente_i\}_{i\in \{0,\cdots, \Delta_v\}}$.
\State Set $\ele_0\gets \top$; for any $ i\in[\Delta_v]$, set $\sent_i\gets 0$, $\rid_i\gets\highlight{2^L-1}$, $\sente_i\gets 0$.
\State $\pieces \gets \ceillb$, $\eachpiece \gets \lceil (\log \ceillb) /B\rceil$. \Comment{End of initialization.}
\For{(each iteration containing $\Theta(\frac{\log{(L/B)}}{B})$ rounds)}
	\State $l\gets\sent_v+1$, $r\gets\min(\sent_v+\eachpiece,\pieces)$.
	\For{(each incident edge with label $i\in[\Delta_v]$)}
		\State $\ischild_i\gets\mathbb{I}[\parent_v==i]$.
		\State Send $\langle \res_v, \build_v, \rid_v[l,\cdots,r], \sent_v, \ischild_i, \flagnew_v, \cnt_v, \ele_v \rangle$ through edge $i$.
	\EndFor
	\State $\sent_v\gets r$.
	\If{($\res_v\neq \perp$)}
		Return $\res_v$ as final result. \Comment{Termination.}
	\EndIf
	\State Let {\scriptsize $m_i=\langle \res_i, \build_i, \rid'_i, \sent'_i, \ischild_i, \flagnew_i, \cnt_i, \ele_i' \rangle$} be the message received via edge {$i\in [\Delta_v]$}.
	\For{(each edge $i\in [\Delta_v]$)}
		\State {\scriptsize $l\gets\sent_i'+1$, $r\gets\min(\sent_i'+\eachpiece,\pieces)$, $\sent_i\gets r$, $\rid_i[l,\cdots, r]\gets\rid'_i$, fill $\rid_i[r+1,\cdots, M]$ with 1.}\label{line:update-rid-i}
		\If{($\res_i\neq \perp$)}
			$\res_v\gets \res_i$.
		\EndIf
		\State $\build_v \gets \build_v \wedge \build_i$.
	\EndFor
\If{($\build_v==true$)}
	Execute Procedure \textsc{BFS-Tree-Building} for large tokens.
\Else
	\xspace Execute Procedure \textsc{Token-Collision-Detection} for large tokens.
\EndIf
\EndFor
\end{algorithmic}
\end{small}
\hrule
\vspace{2ex}
\caption{Pseudocode of the deterministic token collision algorithm for large tokens.}\label{fig:alg-main-alt}
\vspace{-4ex}
\end{figure}

\begin{figure}[t]
\hrule
\vspace{1ex}
{Procedure \textsc{BFS-Tree-Building} executed at node $v$ for large tokens.}
\vspace{1ex}
\hrule
\begin{small}
\begin{algorithmic}[1]
\State $ID_v\gets \{\rid_i \mid  i\in [\Delta_v]\}$.
\If{($\min\{ID_v\}<\rid_v$)}
	\If{($\parent_v \neq \perp$ \textbf{and} $ \rid_{\parent_v} == \min\{ID_v\} $)}
		$j \gets \parent_v$.
	\Else
		\xspace Let $j\in [\Delta_v]$ be one edge label satisfying $\rid_j==\min\{ID_v\}$.
	\EndIf
	\State $\rid_v\gets \rid_j$, $\sent_v\gets \sent_j'$, $\parent_v\gets j$, $\flagnew_v\gets false$. \Comment{Notice that $\flagnew_v$ is reset to $false$.}
\ElsIf{($\max\{ID_v\}==\rid_v$ \textbf{and} $\sent_v==\pieces$ \textbf{and} ($\forall i\in [\Delta_v]$, $\sent_i==\pieces$))}
	\State $\child_v\gets \{i\mid i\in [\Delta_v] \text{ and }\ischild_i==true\}$.
	\If{(($\forall i\in \child_v$, $\flagnew_i==true$) \textbf{or} $\child_v==\emptyset$)}
		$\flagnew_v\gets true$.
	\EndIf
\EndIf
\If{($\parent_v == \perp$ \textbf{and} $\flagnew_v == true$)}
	$\build_v\gets false$.
\EndIf
\end{algorithmic}
\end{small}
\hrule
\vspace{3ex}
\hrule
\vspace{1ex}
{Procedure \textsc{Token-Collision-Detection} executed at node $v$ for large tokens.}
\vspace{1ex}
\hrule
\begin{small}
\begin{algorithmic}[1]
\State $\child_v\gets \{i\mid i\in [\Delta_v] \text{ and } \build_i==false \text{ and } \ischild_i== true \text{ and } \sent_i==\pieces \text{ and } \rid_i==\rid_v\}$.
\For{($i\in \child_v$ \textbf{and} $\ele_i'\notin\{\top, \perp\}$)}
	\State $\ele_i[\sente_i+1,\cdots, \min(\sente_i+\eachpiece, \pieces)] \gets \ele_i'$, $\sente_i\gets \min(\sente_i+\eachpiece, \pieces)$.
	\If{ ($\sente_i==\pieces$)}
		Append $\ele_i$ to $\bm{x}^v$ and set $\sente_i\gets 0$.
	\EndIf
\EndFor
\If{(($\forall i\in [\Delta_v]$, $\build_i==false$) \textbf{and} $\sent_v==\pieces$)}
	\If{(($\forall i\in \child_v$, $\cnt_i\neq \perp$) \textbf{or} $\child_v==\emptyset$)}
		$\cnt_v\gets 1+\sum_{i\in \child_v} \cnt_i$.
	\EndIf
	\If{($\parent_v\neq \perp$)}
		\If{($\sente_0\neq 0$)}
			\State  $l\gets \sente_0+1$, $r\gets \min(\sente_0+\eachpiece, \pieces)$,  $\ele_v\gets \ele_0[l,\cdots, r]$, $\sente_0\gets r \bmod \pieces$.
		\ElsIf{($|\bm{x}^v|>0$)}
			\State Eject one token from $\bm{x}^v$ and let that token be $\ele_0$.
			\State $\ele_v\gets \ele_0[1,\cdots, \eachpiece]$, $\sente_0\gets \eachpiece$.
		\ElsIf{(($\forall i\in \child_v$, $\ele_i==\perp$) \textbf{or} $\child_v==\emptyset$)}
			\xspace $\ele_v\gets \perp$.
		\Else
			\xspace $\ele_v\gets \top$.
		\EndIf
	\ElsIf{($\cnt_v\neq \perp$ \textbf{and} (($\forall i\in \child_v$, $\ele_i== \perp$) \textbf{or} $\child_v==\emptyset$))}
		\If{(know value of $n$ \textbf{and} $\cnt_v== n$ \textbf{and} no token collision in $\bm{x}^v$)}
			$\res_v\gets true$.
		\ElsIf{(know value of $k$ \textbf{and} $|\bm{x}^v|==k$ \textbf{and} no token collision in $\bm{x}^v$)}
			\xspace $\res_v\gets true$.
		\Else
			\xspace $\res_v\gets false$.
		\EndIf
	\EndIf
\EndIf
\end{algorithmic}
\end{small}
\hrule
\vspace{2ex}
\caption{Procedure \textsc{BFS-Tree-Building} and \textsc{Token-Collision-Detection} for large tokens.}\label{fig:alg-procedures-alt}
\vspace{-4ex}
\end{figure}

\section{Omitted Proofs for the Deterministic Algorithm}\label{sec-app:det-alg-proof}

\begin{proof}[\underline{Proof of \Cref{lem:rid-path-increase}}]
We prove by induction on round number that $\rid(v)\geq\rid(u)$ holds for any node $v$ at the end of any round $i$, where $u$ is the parent of $v$ by the end of round $i$.

After initiation, every node $v$ has no parent and the claim holds.

Assume the claim holds for round $i$, we now consider round $i+1$. There are two cases.

In the first case, $\rid(v)$ reduces in round $i+1$. In such case, assume $u$ is the parent of $v$ by the end of round $i+1$, such $u$ must exist. Then, the value of $\rid(v)$ by the end of round $i+1$ is the same as the value of $\rid(u)$ by the end of round $i$, and the value of $\rid(u)$ by the end of round $i+1$ is no more than the value of $\rid(u)$ by the end of round $i$. So the inductive step holds in this case.

In the second case, $\rid(v)$ remains unchanged in round $i+1$. Assume $u$ is parent of node $v$ by the end of round $i$, we know $u$ still is the parent of $v$ by the end of round $i+1$. By the end of round $i$, due to the induction hypothesis, $\rid(v)\geq\rid(u)$. In round $i+1$, value of $\rid(u)$ will not increase. So the inductive step also holds in this case, and this completes the proof of the claim.

Therefore, for any edge $(v,u)$ in any directed path, $\rid(v)\geq\rid(u)$.
\end{proof}

\begin{proof}[\underline{Proof of \Cref{lem:directed-forest}}]
To prove the lemma, the key is to show there are no directed cycles in the identifier-induced graph. Indeed, if we assume there are no directed cycles, then every weakly connected component must be a tree as each node has at most one parent. Consider an arbitrary component, if we take edge directions into consideration, there must exist some node $r$ that has no parent (otherwise the component would contain a directed cycle). This node $r$ is a root of the component. Moreover, such root must be unique for each component. Again, this is due to the fact that in our algorithm each node has at most one parent.

What remains is to show there are no directed cycles in the identifier-induced graph, and we do this by an induction on round number. The induction basis is trivial: after initiation, every node $v$ has no parent and forms a single-node tree.

Assume the claim holds for round $i$, consider the inductive step for round $i+1$.

For the sake of contradiction, assume there exists a directed cycle $v_1, v_2, \cdots, v_t$ with $v_t=v_1$ at the end of round $i+1$. If all nodes in the cycle do not change parent in round $i+1$, then the cycle already exists by the end of round $i$, which contradicts the induction hypothesis. Hence, assume $v_{\hat{j}}$ is the first node in the cycle that changes parent in round $i+1$. Recall that that in our algorithm, a node $v$ changes its parent to some neighbor $u$ in round $i+1$ only if $\rid(v)>\rid(u)$ by the end of round $i$. Therefore, $\rid(v_{\hat{j}})>\rid(v_{\hat{j}+1})$ by the end of round $i$. On the other hand, due to \Cref{lem:rid-path-increase}, $\rid(v_1)\geq\rid(v_2)\geq\cdots\geq\rid(v_{\hat{j}})$ by the end of round $i$. We continue to consider the remaining nodes in the cycle. Each time some node $v_{\tilde{j}}$ changes its parent in round $i+1$, we know $\rid(v_{\tilde{j}})>\rid(v_{\tilde{j}+1})$ by the end of round $i$. Eventually, this implies that by the end of round $i$, it holds that $\rid(v_1)>\rid(v_t)$. But $v_1$ and $v_t$ are the same node, resulting in a contradiction. Therefore, the claim that there are no directed cycles in the identifier-induced graph is proved.
\end{proof}

\begin{proof}[\underline{Proof of \Cref{lem:subtree}}]
Consider an arbitrary node $r$, call the subtree rooted at node $r$ in the identifier-induced graph as $T$. Clearly, any induced subgraph of $T$ is a forest containing one or more trees. By definition, $r$ is included in the considered induced subgraph. Therefore, to prove the lemma, we only need to show that for any node $v$ in $T$ having an identical identifier with $r$, there is a path connecting $v$ and $r$ in the induced subgraph. To see this, notice that in $T$ there is a unique path connecting $v$ to $r$, as $T$ is a tree rooted at $r$. Consider the directed path from $v$ to $r$, by \Cref{lem:rid-path-increase}, the identifiers of the nodes are non-increasing along the path. But $v$ and $r$ have identical identifiers, hence all nodes on this path have identical identifiers as $r$. That is, all nodes and edges on this path will be included in the considered induced subgraph. As a result, there is a path connecting $v$ and $r$ in the induced subgraph, as required.
\end{proof}

\begin{proof}[\underline{Proof of \Cref{lem:tree-of-dinstinct}}]
Throughout the proof, assume there are no token collisions. For each node $v$, let $\vid_v$ denote the minimum token that $v$ received as input. Let $v_{\min}$ denote the unique node having the smallest input token. For any two nodes $u,v\in V$, let $\dist(u,v)$ denote the distance between $u$ and $v$ in the network graph $G$. Define $d=\max_{v\in V}\dist(v,v_{\min})$. For any node $v$, let $d_v$ denote the distance between node $v$ and the nearest node $u$ with $\vid_u$ smaller than $v$. That is, $d_v=\min_{u\in V,\vid_u<\vid_v}\dist(u,v)$. We set $d_{v_{\min}}=+\infty$. We claim:
\begin{enumerate}
	\item \label{item:no-termination} No node quits the BFS-tree building procedure within $d$ rounds. Formally, for any $0\leq i\leq d$, each node $v$ has $\build_v=true$ by the end of round $i$.
	
	\item \label{item:each-node-set-rid} For any node $v$, any $0\leq i\leq d$, let node $\hat{u}$ be the unique node that has minimum $\vid$ among all nodes $u$ with $\dist(u,v)\leq i$. At the end of round $i$, we have $\rid_v=\vid_{\hat{u}}$. Formally, at the end of round $i$, it holds that $\rid_v=\min_{u\in V, \dist(u,v)\leq i}\vid_u$.
	
	\item \label{item:root-tree} For any node $v$, any $0\leq i\leq d$, at the end of round $i$, one of the following cases holds.
	\begin{itemize}
		\item Case I: $d_v>2i$. The nodes with distance at most $i$ to $v$ form a height-$i$ tree rooted at $v$ in the identifier-induced graph. Moreover, for any node $u$ with distance $i$ to $v$, for any node $w$ on the directed path from $u$ to $v$ in the rooted tree, it holds that $\flagnew_w=false$.
		\item Case II: $i<d_v\leq 2i$. There is a tree rooted at $v$ in the identifier-induced graph. Let node $w$ denote the unique node with distance $d_v$ to node $v$ that has the smallest $\vid$. There exists a node $u$ with $\dist(u,v)=d_v-i-1$ and $\dist(u,w)=i+1$ such that $\flagnew$ is $false$ for all nodes along the length-$(d_v-i-1)$ directed path from $u$ to $v$ in the identifier-induced graph.
		\item Case III: $d_v\leq i$. Node $v$ has a parent in the identifier-induced graph.
	\end{itemize}
\end{enumerate}

We prove the above claim by induction on rounds. For the base case $i=0$, which is immediately after the initialization, each node $v$ sets $\build_v$ to $true$, sets $\flagnew_v$ to $false$ and sets $\rid_v$ to $\min\bm{x}^v$. It is easy to verify that all three guarantees are satisfied.

Assume the claim holds by the end of round $i$ where $0\leq i<d$, we consider round $i+1$. By \Cref{item:no-termination} of the induction hypothesis, no node sets $\build$ to $false$ at the end of round $i$. Thus, all nodes execute the BFS-tree building procedure in round $i+1$.

First consider \Cref{item:each-node-set-rid}. By \Cref{item:each-node-set-rid} of the induction hypothesis, $\rid_v=\min_{u\in V, \dist(u,v)\leq i}\vid_u$ for any node $v$ at the end of round $i$. According to the algorithm, at the end of round $i+1$, every node $v$ sets $\rid_v$ to be the minimum $\rid$ (at the end of round $i$) among its inclusive $1$-hop neighborhood, which can be verified to be $\min_{u\in V, \dist(u,v)\leq i+1}\vid_u$.

Then, we prove the inductive step for \Cref{item:root-tree}. Fixed a node $v$, consider following scenarios.
\begin{itemize}
	\item Scenario I: $d_v>2(i+1)$, which corresponds to Case I of \Cref{item:root-tree}. Since $d_v>2(i+1)>2i$, by Case I in \Cref{item:root-tree} of the induction hypothesis, by the end of round $i$, the nodes with distance at most $i$ to $v$ form a height-$i$ tree rooted at $v$ in the identifier-induced graph. Call this tree $T_{v,i}$. Due to \Cref{item:each-node-set-rid} of the induction hypothesis, by the end of round $i$, every node in tree $T_{v,i}$ has $\rid$ equal to $\vid_v$. Moreover, by the end of round $i$, for any node $u$ with distance $i$ to $v$, for any node $w$ on the directed path from $u$ to $v$ in the rooted tree, it holds that $\flagnew_w=false$.
	Now, in round $i+1$, since $d_v>2(i+1)$, for any node $u'$ with $\dist(v,u')\leq i+1$, we know the distance between $u'$ and any node having an $\vid$ smaller than $v$ is larger than $i+1$. Hence, all nodes in $T_{v,i}$ will not change parent in round $i+1$. Moreover, for any node $u'$ with $\dist(v,u')=i+1$, by the end of round $i+1$, its parent will be some node in $T_{v,i}$ that has depth $i$. Hence, by the end of round $i+1$, the nodes with distance at most $i+1$ to $v$ form a height-$(i+1)$ tree rooted at $v$ in the identifier-induced graph. Call this tree $T_{v,i+1}$. Now, consider any node $u'$ with $\dist(v,u')=i+1$, by the end of round $i+1$, there is a directed path from $u'$ to $v$ in $T_{v,i+1}$. For node $u'$, by the end of round $i+1$, $\flagnew_{u'}=false$ since it has changed parent in round $i+1$. For the parent $u$ of $u'$ in this path, by the end of round $i+1$, $\flagnew_{u}=false$ since the $\rid$ received by $u$ from $u'$ in round $i+1$ is different from the one it has (particularly, $\rid_{u'}>\rid_u$ by the end of round $i$). For the remaining nodes on this path, by the end of round $i+1$, its $\flagnew$ will also be $false$ since at least one child (particularly, its child on the path) has passed a $\flagnew$ with $false$ value to it in round $i+1$. Therefore, for any node $u'$ with $\dist(v,u')=i+1$, by the end of round $i+1$, for any node $w'$ on the directed path from $u'$ to $v$ in the rooted tree, it holds that $\flagnew_{w'}=false$.
	
	\item Scenario II: $d_v=2(i+1)$, which corresponds to Case II of \Cref{item:root-tree}. Consider a shortest path (in $G$) between $v$ and $w$, call it $\texttt{path}(v,w)$. Consider the node $u$ with $\dist(v,u)=i+1$ and $\dist(u,w)=i+1$ on that path. Let $u'$ be the neighbor of $u$ on $\texttt{path}(v,w)$ with $\dist(u',v)=i$ and $\dist(u',w)=i+2$. We need to show, by the end of round $i+1$, there is a tree rooted at $v$ in the identifier-induced graph. Moreover, for all nodes along the length-$i$ directed path from $u'$ to $v$ in the identifier-induced graph, their $\flagnew$ value is $false$. To see this, first notice that, by the end of round $i$, by Case I in \Cref{item:root-tree} of the induction hypothesis, all nodes with distance at most $i$ from $v$ form a tree rooted at $v$ in the identifier-induced graph, call this tree $T_{v,i}$; moreover, for all nodes along the length-$i$ directed path from $u'$ to $v$ in $T_{v,i}$, their $\flagnew$ value is $false$. Now, in round $i+1$, since $d_v=2(i+1)$, all nodes in $T_{v,i}$ will not change parent and will still form a tree rooted at $v$. For each node $w'\neq u'$ on the length-$i$ directed path from $u'$ to $v$ in $T_{v,i}$, it holds that $\flagnew_{w'}=false$ by the end of round $i+1$ since at least one child (particularly, its child on the path) has passed a $\flagnew$ with $false$ value to it in round $i+1$. As for $u'$, notice that by the end of round $i$, it holds that $\rid_u>\rid_{u'}=\vid_v$ since $d_v=2(i+1)$. As a result, by the end of round $i+1$, we also have $\flagnew_{u'}=false$ since the $\rid$ received by $u'$ from $u$ in round $i+1$ is different from the one it has.
	
	\item Scenario III: $d_v=2i+1$, which also corresponds to Case II of \Cref{item:root-tree}. Consider a shortest path (in $G$) between $v$ and $w$, call it $\texttt{path}(v,w)$. Consider the node $u$ with $\dist(v,u)=i$ and $\dist(u,w)=i+1$ on that path. Let $u'$ be the neighbor of $u$ on $\texttt{path}(v,w)$ with $\dist(u',v)=i-1$ and $\dist(u',w)=i+2$. We need to show, by the end of round $i+1$, there is a tree rooted at $v$ in the identifier-induced graph. Moreover, for all nodes along the length-$(i-1)$ directed path from $u'$ to $v$ in the identifier-induced graph, their $\flagnew$ value is $false$. To see this, first notice that, by the end of round $i$, by Case I in \Cref{item:root-tree} of the induction hypothesis, all nodes with distance at most $i$ from $v$ form a tree rooted at $v$ in the identifier-induced graph, call this tree $T_{v,i}$; moreover, for all nodes along the length-$i$ directed path from $u$ to $v$ in $T_{v,i}$, their $\flagnew$ value is $false$. Now, in round $i+1$, since $d_v=2i+1$, all nodes in $T_{v,i}$ with depth at most $i-1$ will not change parent and will still form a tree rooted at $v$. Moreover, for each node $w'$ on the length-$(i-1)$ directed path from $u'$ to $v$ in $T_{v,i}$, it holds that $\flagnew_{w'}=false$ by the end of round $i+1$ since at least one child (particularly, its child on the path) has passed a $\flagnew$ with $false$ value to it in round $i+1$.
	
	\item Scenario IV: $i+1< d_v\leq 2i$, which again corresponds to Case II of \Cref{item:root-tree}. Consider a shortest path (in $G$) between $v$ and $w$, call it $\texttt{path}(v,w)$. Consider the node $u$ with $\dist(v,u)=d_v-i-1$ and $\dist(u,w)=i+1$ on that path. Let $u'$ be the neighbor of $u$ on $\texttt{path}(v,w)$ with $\dist(u',v)=d_v-i-2$ and $\dist(u',w)=i+2$. We need to show, by the end of round $i+1$, there is a tree rooted at $v$ in the identifier-induced graph. Moreover, for all nodes along the length-$(d_v-i-2)$ directed path from $u'$ to $v$ in the identifier-induced graph, their $\flagnew$ value is $false$. To see this, first notice that, by the end of round $i$, by Case II in \Cref{item:root-tree} of the induction hypothesis, there is a tree rooted at $v$ in the identifier-induced graph, call this tree $T_{v,i}$; moreover, for all nodes along the length-$(d_v-i-1)$ directed path from $u$ to $v$ in $T_{v,i}$, their $\flagnew$ value is $false$. Now, in round $i+1$, since $d_v>i+1$, node $v$ will not change parent, implying there is still a tree rooted at $v$ by the end of round $i+1$. Moreover, for each node $w'$ on the length-$(d_v-i-2)$ directed path from $u'$ to $v$ in $T_{v,i}$, it holds that $\flagnew_{w'}=false$ by the end of round $i+1$ since at least one child (particularly, its child on the path) has passed a $\flagnew$ with $false$ value to it in round $i+1$.
	
	\item Scenario V: $d_v\leq i+1$, which corresponds to Case III of \Cref{item:root-tree}. Let node $\hat{u}$ be the unique node that has minimum $\vid$ among all nodes $u$ with $\dist(u,v)\leq i+1$. Since $d_v\leq i+1$, by \Cref{item:each-node-set-rid} which is already proved for round $i+1$, node $v$ set its $\rid$ to $\vid_{\hat{u}}$ at the end of round $i+1$. Thus, node $v$ has a parent at the end of round $i+1$.
\end{itemize}

Lastly, we prove the inductive step of \Cref{item:no-termination}. Since no node sets $\build$ to $false$ at the end of round $i$, every node will not receive a termination signal from its neighbors in round $i+1$. Thus, at the end of round $i+1$, any node with a parent will not set $\build$ to $false$ by itself. For a node $v$ without a parent, according to \Cref{item:root-tree} which is already proved for round $i+1$, it has $d_v>i+1$ and $\flagnew_v=false$ at the end of round $i+1$. Thus, a node without a parent will not set $\build$ to $false$ at the end of round $i+1$. This completes the proof of the inductive step.

At this point, the lemma can be easily proved. \highlightblue{By \Cref{item:no-termination-2}, no node will quite the \textsc{BFS-Tree-Building} procedure within $d$ rounds.} By \Cref{item:each-node-set-rid}, at the end of round $d$, all nodes have identical $\rid$, which is $\vid_{v_{min}}$. By Case I of \Cref{item:root-tree}, at the end of round $d$, all nodes form a single tree rooted at $v_{min}$ in the identifier-induced graph. Moreover, no node will ever change parent or $\rid$ later.
\end{proof}

\begin{proof}[\underline{Proof of \Cref{lem:cnt-collect}}]
Notice that in our algorithm, only a root node in the identifier-induced graph can update its $\res$ within procedure \textsc{Token-Collision-Detection}, so $v$ must be a root node in the identifier-induced graph by the end of round $i$. Let $T_{v,i}$ be the identifier-induced subtree rooted at $v$ by the end of round $i$. We need to show $\cnt_v$ equals the size of $T_{v,i}$ by the end of round $i$.

To prove the above result, we claim: for any node $u$, if $i_u$ is the first round in which $u$ updates $\cnt_u$ to some non-$\perp$ value, then $\cnt_u$ equals the size of $T_{u,i_u}$ by the end of round $i_u$, where $T_{u,i_u}$ is the identifier-induced subtree rooted at $u$ by the end of round $i_u$. Moreover, by the end of any round $i'>i_u$, $\cnt_u$ remains unchanged and $T_{u,i'}$ is identical to $T_{u,i_u}$. We will prove this claim by induction on round number.

We begin with the base case, let $\hat{i}$ be the first round in which some node $u$ updates $\cnt_u$ to some non-$\perp$ value. In this case, note that in our algorithm, any node $w$ only updates $\cnt_w$ to some non-$\perp$ value if all of its children in the identifier-induced subtree has already obtained a non-$\perp$ valued $\cnt$, or $w$ has no children in the identifier-induced subtree. Since $\hat{i}$ is the first round in which some node updates $\cnt$ to some non-$\perp$ value, we know $u$ has no children in $T_{u,\hat{i}}$, meaning $T_{u,\hat{i}}$ is a single-node tree and $\cnt_u=1$ by the end of round $\hat{i}$, as required. On the other hand, in our algorithm, any node $u$ only executes procedure \textsc{Token-Collision-Detection} after all its neighbors stop building BFS-trees---more precisely, after all its neighbors set $\build=false$. Moreover, once a node stops building BFS-tree, the node will not update its parent. Therefore, all neighbors of $u$ will not change parent pointer in any round $i>\hat{i}$. As a result, by the end of any round $i>\hat{i}$, $\cnt_u$ remains unchanged and $T_{u,i}$ is identical to $T_{u,\hat{i}}$. By now we have proved the base case.

Assume the claim holds for any round up to $\tilde{i}-1\geq\hat{i}$, now consider round $\tilde{i}$. If in round $\tilde{i}$ no node changes $\cnt$ from $\perp$ to some integer, then the induction hypothesis already implies the inductive step. So assume some node $u$ updates $\cnt_u$ from $\perp$ to some integer in this round. If $u$ has no children in $T_{u,\tilde{i}}$, then the inductive step can be proved by a similar argument as in the base case. So assume $u$ has at least one child in $T_{u,\tilde{i}}$. Consider an arbitrary child $w$ of $u$ in $T_{u,\tilde{i}}$. Node $u$ must have received a $\cnt_w\neq\perp$ from $w$ in round $\tilde{i}$. This implies, if $i_w$ is the first round in which $w$ updates $\cnt_w$ to some non-$\perp$ value, then $i_w<\tilde{i}$. By the induction hypothesis, the $\cnt_w$ node $w$ passed to $u$ in round $\tilde{i}$ is identical to the value it obtained in round $i_w$, which is $|T_{w,i_w}|$. Moreover, $T_{w,i_w}$ and $T_{w,\tilde{i}}$ are identical. Therefore, by the end of round $i$, we know $\cnt_u=|T_{u,\tilde{i}}|$, as required. On the other hand, in our algorithm, any node $u$ only executes procedure \textsc{Token-Collision-Detection} after all its neighbors stop building BFS-trees---more precisely, after all its neighbors set $\build=false$. Moreover, once a node stops building BFS-tree, the node will not update its parent. Therefore, all neighbors of $u$ will not change parent pointer in any round $i>\tilde{i}$. Therefore, by the end of any round $i>\tilde{i}$, $\cnt_u$ remains unchanged and $T_{u,i}$ is identical to $T_{u,\tilde{i}}$. This completes the proof for the inductive step.

With the above claim, the lemma is easy to obtain. Assume $i_v$ is the first round in which $v$ updates $\cnt_v$ to some non-$\perp$ value, then $\cnt_v$ equals the size of $T_{v,i_v}$ by the end of round $i_v$. Later by the end of round $i$, when $v$ updates $\res$ to some non-$\perp$ value in procedure \textsc{Token-Collision-Detection}, the $\cnt_v$ value remains unchanged and is $|T_{v,i_v}|=|T_{v,i}|$.
\end{proof}

\begin{proof}[\underline{Proof of \Cref{lem:token-collect}}]
The high-level proof strategy is similar to that of \Cref{lem:cnt-collect}. Notice that in our algorithm, only a root node in the identifier-induced graph can update its $\res$ within \textsc{Token-Collision-Detection}, so $v$ must be a root node in the identifier-induced graph by the end of round $i$. Let $T_{v,i}$ be the identifier-induced subtree rooted at $v$ by the end of round $i$. We need to show that $v$ collects each token owned by the nodes in $T_{v,i}$ exactly once by the end of round $i$.

To that end, we argue the following claim. For any node $u$, if $i_u$ is the first round in which $u$ runs \textsc{Token-Collision-Detection}, and obtains a $\perp$-valued $\ele$ from each of its children in the identifier-induced subtree or $u$ has no children, then:
\begin{enumerate}
	\item\label{item:token-collect} Node $u$ has collected each token owned by the nodes in $T_{u,i_u}$ once by the end of round $i_u$, where $T_{u,i_u}$ is the identifier-induced subtree rooted at $u$ by the end of round $i_u$.
	\item\label{item:token-collect-persistent} No new tokens will be passed to $u$ in any round $i'>i_u$, and $T_{u,i'}$ is identical to $T_{u,i_u}$.
	\item\label{item:token-collect-prev} If node $u$ has sent an $\ele$ containing a token to its parent in some round $i''\leq i_u$, then that token is owned by some node in $T_{u,i_u}$.
	\item\label{item:token-collect-finish} If $i''_u\geq i_u$ is the first round in which $u$ has a parent and sets $\ele$ to $\perp$, then $u$ has sent each token in $T_{u,i_u}$ to that parent exactly once by the end of round $i''_u$. Moreover, $u$ will not send any more tokens to that parent later.
\end{enumerate}
We will prove this claim by induction on round number.

We begin with the base case, let $\hat{i}$ be the first round in which some node $u$ runs \textsc{Token-Collision-Detection}, and obtains a $\perp$-valued $\ele$ from each of its children in the identifier-induced subtree or $u$ has no children in the identifier-induced subtree. Notice that in our algorithm, each node's $\ele$ is initialized to $\top$, which implies it can only send a $\perp$-valued $\ele$ to its parent after itself has already set $\ele$ to $\perp$. Therefore, in round $\hat{i}$, node $u$ has no children in $T_{u,\hat{i}}$, meaning $T_{u,\hat{i}}$ is a single-node tree. We now proceed to prove the four items in the claim.

First, for \Cref{item:token-collect}, since $T_{u,\hat{i}}$ is a single-node tree, it trivially holds.

Then, for \Cref{item:token-collect-persistent}, notice that in our algorithm, any node $u$ only executes \textsc{Token-Collision-Detection} after all its neighbors stop building BFS-trees---more precisely, after all its neighbors set $\build=false$. Moreover, once a node stops building BFS-tree, the node will not update its parent. Hence, once a node starts executing \textsc{Token-Collision-Detection}, the topology of its one-hop neighborhood is fixed. In particular, all neighbors of $u$ will not change parent pointer in any round $i'>\hat{i}$. As a result, by the end of any round $i'>\hat{i}$, no new tokens will be passed to $u$ in round $i'$, and $T_{u,i'}$ is identical to $T_{u,\hat{i}}$.

Next, for \Cref{item:token-collect-prev}, recall that we just argued that once a node starts executing \textsc{Token-Collision-Detection}, the topology of its one-hop neighborhood is fixed. Since $T_{u,\hat{i}}$ is a single-node tree and $u$ only starts sending tokens to parent after it starts executing \textsc{Token-Collision-Detection}, any token $u$ sent in round $i''\leq\hat{i}$ is owned by some node in $T_{u,i_u}$---particularly, itself.

Lastly, consider \Cref{item:token-collect-finish}. Again, recall that once a node starts executing \textsc{Token-Collision-Detection}, the topology of its one-hop neighborhood is fixed. Since $T_{u,\hat{i}}$ is a single-node tree and $u$ only starts sending tokens to parent after it starts executing \textsc{Token-Collision-Detection}, and since $u$ only sets $\ele$ to $\perp$ after it has sent all tokens, \Cref{item:token-collect-finish} holds.

This completes the proof for the base case.

Assume the claim holds for any round up to $\tilde{i}-1\geq\hat{i}$, now consider round $\tilde{i}$. If, for some node $u$, this is the first round that it discovers it has no children in $T_{u,\tilde{i}}$ while $u$ running \textsc{Token-Collision-Detection}, then the inductive step can be proved for $u$ by a similar argument as in the base case. Otherwise, assume for some node $u$, this is the first round that it discovers every child of it in $T_{u,\tilde{i}}$ has sent a $\perp$-valued $\ele$ to it while $u$ running \textsc{Token-Collision-Detection}. We now prove the four items in the claim.

We begin with \Cref{item:token-collect}. Consider an arbitrary child $w$ of $u$ in $T_{u,\tilde{i}}$. Let $i_w<\tilde{i}$ be the first round in which $w$ runs \textsc{Token-Collision-Detection}, and obtains a $\perp$-valued $\ele$ from each of its children in $T_{w,i_w}$ or $w$ has no children in $T_{w,i_w}$. By \Cref{item:token-collect-finish} of the induction hypothesis, $u$ has collected each token in $T_{w,i_w}$ exactly once by the end of round $\tilde{i}$. By \Cref{item:token-collect-persistent} of the induction hypothesis, $T_{w,\tilde{i}}=T_{w,i_w}$. Hence, $u$ has collected each token owned in $T_{u,\tilde{i}}$ exactly once by the end of round $\tilde{i}$.

Then, for \Cref{item:token-collect-persistent}, recall that we previously argued that once a node starts executing \textsc{Token-Collision-Detection}, the topology of its one-hop neighborhood is fixed. Therefore, $u$'s neighbors will not change parent pointers in any round $i'>\tilde{i}$. Moreover, by the induction hypothesis, for any child $w$ of $u$ in $T_{u,\tilde{i}}$, for any round $i'>\tilde{i}$, by \Cref{item:token-collect-persistent} of the induction hypothesis, $T_{w,i'}=T_{w,\tilde{i}}$. Hence, for any round $i'>\tilde{i}$, $T_{u,i'}=T_{u,\tilde{i}}$. On the other hand, for any child $w$ of $u$ in $T_{u,\tilde{i}}$, for any round $i'>\tilde{i}$, by \Cref{item:token-collect-finish} of the induction hypothesis, no tokens will be passed to $u$ by $w$. Hence, for any round $i'>\tilde{i}$, no tokens will be passed to $u$.

Next, we consider \Cref{item:token-collect-prev}. Assume $u$ has sent an $\ele$ containing a token to its parent in some round $i''\leq\tilde{i}$. If that token is owned by $u$, then the claim in the item trivially holds. So, assume that token is forwarded by some neighbor $w$ in some round $i''_w<\tilde{i}$. Since the one-hop neighborhood of $u$ is fixed once it starts executing \textsc{Token-Collision-Detection}, $w$ is a child of $u$ in $T_{u,\tilde{i}}$. Moreover, it is easy to verify that the token forwarded by $w$ is owned by some node in $T_{w,\tilde{i}}$.

Lastly, we consider \Cref{item:token-collect-finish}. By \Cref{item:token-collect-prev} of the inductive step which we have proved, if $u$ has sent a token to its parent in some round before (and including) $\tilde{i}$, then that token is owned by some node in $T_{u,\tilde{i}}$. By \Cref{item:token-collect} of the inductive step which we have proved, $u$ has collected each token owned by the nodes in $T_{u,\tilde{i}}$ exactly once by the end of round $\tilde{i}$. Notice that by our algorithm description, If $i''_u$ is the first round in which $u$ has a parent and sets $\ele$ to $\perp$, then $i''_u\geq\tilde{i}$. Moreover, our algorithm will only sets $\ele$ to $\perp$ if it has sent each token in the token list to its parent. Therefore, $u$ has sent each token in $T_{u,\tilde{i}}$ to its parent exactly once by the end of round $i''_u$. Moreover, by \Cref{item:token-collect-persistent} of the induction hypothesis, every child of $u$ will not send tokens to $u$ in $i''_u$ or later, so $u$ will not send any more tokens to its parent in round $i''_u+1$ or later.

Now that the claim is proved, the lemma is easy to obtain. Specifically, assume $i_v$ is the first round in which $v$ runs \textsc{Token-Collision-Detection}, and obtains a $\perp$-valued $\ele$ from each of its children in $T_{v,i_v}$ or $v$ has no children in $T_{v,i_v}$. Then, by \Cref{item:token-collect}, $v$ has collected each token owned by the nodes in $T_{v,i_v}$ exactly once by the end of round $i_v$. Moreover, by \Cref{item:token-collect-persistent}, no tokens will be passed to $v$ in later rounds and the identifier-induced subtree rooted at $v$ remains unchanged. Therefore, later by the end of some round $i\geq i_v$, when $v$ updates $\res$ to some non-$\perp$ value, it has collected each token owned by the nodes in $T_{v,i}$ exactly once.
\end{proof}

\begin{proof}[\underline{Proof of \Cref{lem:comp-min-subtree-D}}]
By algorithm description, node $v$ will never have a parent because it holds a global minimum token, thus $v$ is always a root in the identifier-induced graph.

Consider an arbitrary round $i$, assume node $u$ is in the identifier-induced subtree rooted at $v$ by the end of round $i$. Consider the path from $u$ to $v$ in this identifier-induced subtree, we claim this path must be a shortest path connecting $u$ and $v$ in the network graph $G$. Otherwise, it means that on every shortest path connecting $u$ and $v$ in the network graph $G$, there is a node $w$ that quits the BFS-tree building procedure (that is, sets $\build_w=false$) before round $i_w$, where $i_w$ is the distance between $w$ and $v$ in $G$. But in such case, due to the flooding mechanism of a $false$-valued $\build$ variable, before $u$ can receive $\rid_v$, it will set $\build_u$ to $false$, hence will not join the identifier-induced subtree rooted at $v$, resulting in a contradiction. Hence, the path from $u$ to $v$ in the identifier-induced subtree rooted at $v$ must be a shortest path connecting $u$ and $v$ in the network graph $G$. Therefore, the length of this path is $O(D)$.
\end{proof}

\begin{proof}[\underline{Proof of \Cref{lem:comp-all-quit-decide}}]
Recall our algorithm guarantees that if one node generates an $\res\neq\perp$ and then halts, then this $\res$ is broadcast to all other nodes. Hence, all other nodes will halt within $D$ rounds. As a result, to prove the lemma, we show some node will halt within $O(D+k)$ rounds. To this end, we will show that after all nodes quit \textsc{BFS-Tree-Building} which happens within $O(D)$ rounds (by \Cref{lem:comp-all-quit-BFS}), there exists a tree in the identifier-induced graph of height $O(D)$ (by \Cref{lem:comp-min-subtree-D}), and convergecasting tokens inside this tree takes $O(D+k)$ rounds.

In our algorithm, after all nodes have sent $\build=false$ to neighbors, and before any node halts, a node is always sending tokens to its parent if its token list is not empty.

We argue that, after all nodes have sent $\build = false$ to neighbors, if an identifier-induced subtree rooted at node $v$ has height $h_v$ and contains $k_v$ tokens, then $v$ can collect each of the $k_v$ tokens exactly once within $C(h_v+k_v)$ rounds, where $C$ is some positive constant. (We use constant $C$ because every node will send tokens as well as the signal $\ele = \perp$ to suggest that the sending has finished, thus the time complexity is not exactly $h_v+k_v$, but still linear in $h_v+k_v$.) Call the identifier-induced subtree rooted at $v$ as $T_v$. We prove the claim by an induction on $k_v$.

The base case is obvious, if there is only one token ($k_v = 1$) in $T_v$, then convergecast takes at most $C(h_v+1)$ rounds.

Assume the claim holds for $k_v = i$, now consider the case $k_v = i+1$. For the sake of proof, assume tokens are arbitrarily labeled as $t_1,\cdots,t_{i+1}$. For $t_1,\cdots,t_{i}$, by the induction hypothesis, it takes $C(h_v+i)$ rounds to convergecast them. For the additional token $t_{i+1}$, we consider two cases.

In the first case, along the path on which $t_{i+1}$ is sent to the root, no node has tokens that pending to be sent when $t_{i+1}$ arrives (that is, each node's token list is empty when $t_{i+1}$ arrives). As a result, sending $t_{i+1}$ does not affect the sending of $t_1,\cdots, t_{i}$, and sending $t_{i+1}$ itself takes at most $C(h_v+1)$ rounds. Therefore, it takes $\max\{C(h_v+i), C(h_v+1)\}=C(h_v+i)$ rounds in total.

In the second case, along the path where $t_{i+1}$ is sent to the root, there exists some node sending tokens when $t_{i+1}$ arrives. Assume $u$ is one such node. Imagine $u$ takes the strategy that it sends all tokens in $\{t_1,\cdots,t_i\}$ pending in its token list, and then send $t_{i+1}$. Intuitively, $t_{i+1}$ always waits until $u$ is ``free", and ``follow" the last token in $\{t_1,\cdots,t_i\}$ that $u$ intends to send. Hence, $t_{i+1}$ is always $C$ rounds ``behind'' the token it is following. As a result, $t_{i+1}$ does not affect the sending of $t_1,\cdots,t_{i}$, implying tokens $t_1,\cdots,t_{i}$ are sent to the root $v$ within $C(h_v+i)$ rounds by the induction hypothesis. Moreover, $t_{i+1}$ is sent to the root $v$ within $C(h_v+i+1)$ rounds because it is $C$ rounds ``behind'' the last token it ``follow''.

Notice that the above strategy of sending $t_{i+1}$ is equivalent to our algorithm's behavior, as the tokens are arbitrarily labeled.
By now, we have proved by induction that, after all nodes have sent $\build = false$ to neighbors, if an identifier-induced subtree rooted at $v$ has height $h_v$ and contains $k_v$ tokens, then $v$ can collect each token exactly once in the tree in $O(h_v+k_v)$ rounds.

By \Cref{lem:comp-all-quit-BFS}, we know that all nodes have sent $\build = false$ to neighbors in $O(D)$ rounds.
Then, consider a node $v$ which has a global minimum token as input. In the identifier-induced graph, by \Cref{lem:comp-min-subtree-D}, there is a tree rooted at $v$ and the identifier-induced subtree rooted at $v$ has height $O(D)$ and contains at most $k$ tokens. By our above analysis, in $O(D+k)$ rounds, either $v$ collects all tokens in its tree, or some node already halts. Moreover, it is easy to see in $O(D)$ rounds, either $v$ has obtained a count on the size of its tree (that is, $\cnt\neq\perp$), or some node already halts.
At this point, we conclude, within $O(D+k)$ rounds since the start of execution, some node will halt.
\end{proof}

\section{Omitted Proofs for the Generalized Deterministic Algorithm}\label{sec-app:det-alg-proof-gen}

\begin{proof}[\underline{Proof of \Cref{lem:rid-path-increase-gen}}]
We prove by induction on iteration that $\rid_v\geq\rid_u$ holds for any node $v$ at the end of any iteration $i$, where $u$ is the parent of $v$ by the end of iteration $i$.

After initiation, every node $v$ has no parent and the claim holds.

Assume the claim holds for iteration $i$, consider iteration $i+1$. There are two cases.

In the first case, $\rid_v$ reduces in iteration $i+1$. In such case, assume $u$ is the parent of $v$ by the end of iteration $i+1$, such $u$ must exist. Let $\rid'_{u}$ be the identifier prefix $v$ holds for $u$. According to our algorithm, the value of $\rid'_{u}$ by the end of iteration $i+1$ is at least the value of $\rid_{u}$ by the end of iteration $i$. Moreover, the value $\rid_v$ by the end of iteration $i+1$ is updated to the value of $\rid_{u'}$ by the end of iteration $i+1$. Hence, the value of $\rid_v$ by the end of iteration $i+1$ is at least the value of $\rid_u$ by the end of iteration $i$. Notice that the value of $\rid_u$ by the end of iteration $i+1$ is at most the value of $\rid_u$ by the end of iteration $i$, the inductive step holds in this case.

In the second case, $\rid_v$ remains unchanged in iteration $i+1$. Assume $u$ is the parent of $v$ by the end of iteration $i$, we know $u$ still is the parent of $v$ by the end of iteration $i+1$. Due to the induction hypothesis, $\rid_v\geq\rid_u$ by the end of iteration $i$. In iteration $i+1$, the value of $\rid_u$ will not increase, so the inductive step also holds in this case.
\end{proof}

\begin{proof}[\underline{Proof of \Cref{lem:tree-of-dinstinct-gen}}]
Throughout the proof, assume there are no token collisions.

We reuse some notations that are introduced in the proof of \Cref{lem:tree-of-dinstinct}. For each node $v$, let $\vid_v$ denote the minimum token that $v$ received as input. Let $v_{\min}$ denote the unique node having the smallest input token. For any two nodes $u,v\in V$, let $\dist(u,v)$ denote the distance between $u$ and $v$ in the network graph $G$. Define $d=\max_{v\in V}\dist(v,v_{\min})$.

We also need some new notations to facilitate presentation. For any binary string $s$ of length $|s|\leq L$, define $s\uplus\infty$ to be the length-$L$ binary string that begins with $s$ and filled with $1$ thereafter. \highlightblue{Recall that the bandwidth of the network is $B=\Theta(\log{n})$.}  Define constant $\rounds=\lceil\pieces /\eachpiece\rceil$, where $\pieces=\ceillb$ denotes the number of pieces a token needs to be divided into, and $\eachpiece=\lceil(\log{\ceillb})/B\rceil$ denotes the number of pieces that can be transferred in one iteration. (Hence, $\rounds$ denotes the number of iterations required to transfer an entire token.)

For any node $v$, any $0\leq i\leq d+\rounds-1$, any \highlight{$0\leq j\leq i$}, define function $f(v,i,j)$ as follows,
$$
f(v,i,j)=
\begin{cases}
\vid_v & \textrm{if }j=0, \\
\min_{u\in V, \dist(u,v)=j} \vid_u[\leq \min((i-j+1)\cdot \eachpiece, \pieces)]\uplus \infty & \textrm{otherwise}.
\end{cases}
$$
Here, $\vid[\leq x]$ denotes the first $x$ pieces of $\vid$.
\highlight{Intuitively, $f(v,i,j)$ is trying to capture ``considering the nodes that are $j$ hops away from $v$, what is the smallest $\vid$ prefix node $v$ can potentially see regarding these nodes when the algorithm is executed for $i$ iterations''.}
Then, define function $g_{\min}$ in the following way,
\[\highlight{g_{\min}(v,i)}=\min\argmin_{\highlight{0\leq j\leq i}} f(v,i,j).\]
\highlight{Intuitively, $g_{\min}(v,i)$ is trying to capture ``if the algorithm is executed for $i$ iterations, among the smallest $\vid$ prefixes $v$ saw, what is the distance between $v$ and the nearest node that has such an $\vid$ prefix''.}
For any node $v\neq v_{\min}$, let $t_v$ denote the smallest integer such that $g_{\min}(v,t_v) \neq 0$ and let $d_v$ denote $g_{\min}(v,t_v)$.
We further define $t_{v_{\min}}$ and $d_{v_{\min}}$ to be $+\infty$.

We make the following three claims and prove their correctness via induction on rounds.
\begin{enumerate}
	\item\label{item:no-termination-gen}
	No node quits the BFS-tree building procedure within $d+\rounds-1$ iterations. Formally, for any $0\leq i\leq d+\rounds-1$, each node $v$ has $\build_v=true$ by the end of iteration $i$.

	\item\label{item:each-node-set-rid-sent-gen}
	For any node $v$, any $0\leq i\leq d+\rounds-1$, any $j\in [\Delta_v]$, let $u$ denote the node on the other endpoint of $v$'s $j$-th link. \highlight{Upon entering the BFS-tree building procedure in iteration $i$}, $\rid_j$ in the view of $v$ satisfies
	\[\rid_j=\begin{cases}
		\rid_u & \textrm{if }\parent_u\neq \perp, \\
		\rid_u[\leq \min(i\cdot \eachpiece, \pieces)]\uplus \infty & \textrm{otherwise}.
	\end{cases}\]
	Here, $\rid_u$ and $\parent_u$ denote the value of $\rid$ and $\parent$ of $u$ at the end of iteration $i-1$, and $\rid[\leq x]$ denotes the first $x$ pieces of $\rid$.
	Moreover, at the end of iteration $i$, we have
	$$\rid_v=f(v,i,g_{\min}(v,i))\textrm{ and }\sent_v= \min ( (i-g_{\min}(v,i))\cdot \eachpiece,  \pieces).$$

	\item\label{item:each-node-false-gen}
	For any node $v$, any $0\leq i\leq d+\rounds-1$, at the end of iteration $i$, one of the following three cases holds.
	\begin{itemize}
		\item Case I: $i<t_v$ and $\max(i-\rounds+1,0)+i<t_v$. Let $h=\max(i-\rounds+1,0)$. The nodes with distance at most $h$ to $v$ form a \highlightblue{tree rooted at $v$} in the identifier-induced graph. Moreover, for any node $u$ with distance $h$ to $v$, for any node $w$ on the directed path from $u$ to $v$ in the identifier-induced graph, it holds that $\flagnew_w=false$.
		\item Case II: $i<t_v$ and $\max(i-\rounds+1,0)+i\geq t_v$. Let $h=\max(i-\rounds+1,0)$. There is a tree rooted at $v$ in the identifier-induced graph. Let $w$ denote the node with distance $d_v$ to $v$ with the smallest $\vid$. There exists a node $u$ with $\dist(v,u)=t_v-i-1$ and $\dist(u,w)=i-(t_v-d_v)+1$ such that $\flagnew$ is $false$ for all nodes along the length-$(t_v-i-1)$ directed path from $u$ to $v$ in the identifier-induced graph.
		\item Case III: $i\geq t_v$. Node $v$ has a parent. That is, $\parent_v\neq \perp$.
	\end{itemize}
\end{enumerate}

We prove the above claim by induction on rounds.

For the base case $i=0$, which is immediately after initialization, each node $v$ sets $\build_v$ to $true$, sets $\flagnew_v$ to $false$, sets $\rid_j$ to \highlightblue{$2^L-1$} for all $j\in [\Delta_v]$, sets $\sent_v$ to $0$ and sets $\rid_v$ to $\min\bm{x}^v$. It is easy to verify that all three items are satisfied.

Assume the claim holds by the end of iteration $i$ where $0\leq i<d+\rounds-1$, we consider iteration $i+1$. By \Cref{item:no-termination} of the induction hypothesis, no node sets $\build$ to $false$ at the end of iteration $i$. Thus, all nodes execute the BFS-tree building procedure in iteration $i+1$.

First, we focus on \Cref{item:each-node-set-rid-sent-gen}.
For any node $v$, any $j\in [\Delta_v]$, let $u$ denote the node on the other endpoint of $v$'s $j$-th link.
By \Cref{item:each-node-set-rid-sent-gen} of the induction hypothesis, upon entering the BFS-tree building procedure in iteration $i$, $\rid_j$ in the view of $v$ satisfies
\[
\rid_j=
\begin{cases}
	\rid_u & \textrm{if }\parent_u\neq\perp, \\
	\rid_u[\leq \min(i\cdot \eachpiece, \pieces)]\uplus \infty & \textrm{otherwise}.
\end{cases}
\]
Here, $\rid_u$ and $\parent_u$ denote the value of $\rid$ and $\parent$ of $u$ at the end of iteration $i-1$.
According to the algorithm, if $\parent_u=\perp$ at the end of iteration $i$, then $\parent_u=\perp$ at the end of iteration $i-1$, and $u$ sends following pieces of its $\vid$ to $v$ in the $(i+1)$-th iteration and $v$ updates $\rid_j$. Thus, if $\parent_u=\perp$ at the end of iteration $i$, then $\rid_j$ equals to $\rid_u[\leq \min((i+1)\cdot \eachpiece, \pieces)]\uplus \infty$ when $v$ enters the BFS-tree building procedure in iteration $i+1$, as required.
\highlightblue{If $\parent_u\neq \perp$ at the end of iteration $i$, then according to the algorithm, it is easy to verify that, when node $v$ enters the BFS-tree-building procedure in iteration $i+1$, $\rid_j$ in the view of $v$ equals to the value of $\rid_u$ at the end of iteration $i$, as required.}
\highlightblue{On the other hand, by the definition of function $f$ and $g_{\min}$, and the algorithm description, one can also verify that $\rid_v=f(v,i+1,g_{\min}(v,i+1))$ and $\sent_v=\min( (i+1-g_{\min}(v,i+1))\cdot\eachpiece,\pieces)$ are satisfied at the end of iteration $i+1$.}

Then, we prove \Cref{item:each-node-false-gen}. Let $h=\max(i-Q+1,0)$ and $h'=\max(i-Q+2,0)$. For a fixed node $v$, we consider following scenarios.

\begin{itemize}
	\item Scenario I: $i+1<t_v$ and $h'+i+1<t_v$.  Since $h'+i+1<t_v$, it holds that $h+i<t_v$. By Case I in \Cref{item:each-node-false-gen} of the induction hypothesis, at the end of iteration $i$, the nodes with distance at most $h$ to $v$ form a \highlightblue{tree rooted at $v$} in the identifier-induced graph. Moreover, for any node $u$ with distance $h$ to $v$, for any node $w$ on the directed path from $u$ to $v$, it holds that $\flagnew_w=false$. Call this tree $T_{v,i}$. Due to \Cref{item:each-node-set-rid-sent-gen} of the induction hypothesis, by the end of iteration $i$, every node in tree $T_{v,i}$ has $\rid$ equal to $\vid_v$.
	Since $h'+i+1<t_v$, for any node  $u'$ with $\dist(v,u')\leq h'$, at the end of iteration $i$, it holds that $\rid_{u'}\geq \vid_v$. Hence, all nodes in $T_{v,i}$ will not change parent in iteration $i+1$. Moreover, if $h'>0$, for any node $u'$ with $\dist(v,u')=h'$, by the end of iteration $i+1$, its parent will be some node in $T_{v,i}$ that has depth $h$. Hence, by the end of iteration $i+1$, the nodes with distance at most $h'$ to $v$ form a height-$h'$ tree rooted at $v$ in the identifier-induced graph. Call this tree $T_{v,i+1}$. Now, consider any node $u'$ with $\dist(v,u')=h'$, by the end of iteration $i+1$, there is a length-$h'$ directed path from $u'$ to $v$ in $T_{v,i+1}$. For node $u'$, by the end of iteration $i+1$, $\flagnew_{u'}=false$ since \highlight{$\sent_{u'}\neq M$}. For the parent $u$ of $u'$ on this path, by the end of iteration $i+1$, $\flagnew_{u}=false$ since either the $\rid$ received by $u$ from $u'$ in iteration $i+1$ is different from the one it has (particularly, $\rid_{u'}>\rid_u$ by the end of iteration $i$). For the remaining nodes on this path, by the end of iteration $i+1$, its $\flagnew$ will also be $false$ since at least one child (particularly, its child on the path) has passed a $\flagnew$ with $false$ value to it in iteration $i+1$. Therefore, for any node $u'$ with $\dist(v,u')=h'$, by the end of iteration $i+1$, for any node $w'$ on the directed path from $u'$ to $v$ in the rooted tree, it holds that $\flagnew_{w'}=false$.

	\item Scenario II: $i+1<t_v$ and $h'+i+1=t_v$, which corresponds to Case II of \Cref{item:each-node-false-gen}. Consider the node $u$ with $\dist(v,u)=t_v-i-1$ and $\dist(u,w)=i-(t_v-d_v)+1$. Let node $u'$ denote the neighbor of $u$ with $\dist(v,u')=t_v-i-2=h$ and $\dist(u',w)=i-(t_v-d_v)+2$. We need to show, by the end of iteration $i+1$, there is a tree rooted at $v$ in the identifier-induced graph. Moreover, for all nodes along the length-$(t_v-i-2)$ directed path from $u'$ to $v$ in the identifier-induced graph, their $\flagnew$ value is $false$.
	To see this, first notice that, by the end of iteration $i$, by Case I in \Cref{item:each-node-false-gen} of the induction hypothesis, all nodes with distance at most $h$ from $v$ form a tree rooted at $v$ in the identifier-induced graph, call this tree $T_{v,i}$; moreover, for all nodes along the length-$h$ directed path from $u'$ to $v$ in $T_{v,i}$, their $\flagnew$ value is $false$.
	Now, in iteration $i+1$, since $i+1<t_v$ and $h'+i+1=t_v$, all nodes in $T_{v,i}$ will not change parent and will still form a tree rooted at $v$. For each node $w'\neq u'$ on the length-$(t_v-i-2)$ directed path from $u'$ to $v$ in $T_{v,i}$, it holds that $\flagnew_{w'}=false$ by the end of iteration $i+1$ since at least one child (particularly, its child on the path) has passed a $\flagnew$ with $false$ value to it in iteration $i+1$. As for $u'$, notice that by the end of iteration $i$, it holds that either $\rid_u>\rid_{u'}=\vid_v$ or $\sent_{u}\neq M$. As a result, by the end of iteration $i+1$, we also have $\flagnew_{u'}=false$.

	\item Scenario III: $i+1<t_v$ and $h'+i+1=t_v+1$, which also corresponds to Case II of \Cref{item:each-node-false-gen}. Consider the node $u$ with $\dist(v,u)=t_v-i-1=h$ and $\dist(u,w)=i-(t_v-d_v)+1$. Let $u'$ be the neighbor of $u$ with $\dist(u',v)=t_v-i-2=h-1$ and $\dist(u',w)=i-(t_v-d_v)+2$. We need to show, by the end of iteration $i+1$, there is a tree rooted at $v$ in the identifier-induced graph. Moreover, for all nodes along the length-$(t_v-i-2)$ directed path from $u'$ to $v$ in the identifier-induced graph, their $\flagnew$ value is $false$.
	To see this, first notice that, by the end of iteration $i$, by Case I in \Cref{item:root-tree} of the induction hypothesis, all nodes with distance at most $h$ from $v$ form a tree rooted at $v$ in the identifier-induced graph, call this tree $T_{v,i}$; moreover, for all nodes along the length-$h$ directed path from $u$ to $v$ in $T_{v,i}$, their $\flagnew$ value is $false$.
	Now, in iteration $i+1$, since $h'+i+1=t_v+1$, all nodes in $T_{v,i}$ with depth at most $h-1$ will not change parent and will still form a tree rooted at $v$. Moreover, for each node $w'$ on the length-$(h-1)$ directed path from $u'$ to $v$ in $T_{v,i}$, it holds that $\flagnew_{w'}=false$ by the end of iteration $i+1$ since at least one child (particularly, its child on the path) has passed a $\flagnew$ with $false$ value to it in iteration $i+1$.

	\item Scenario IV: $i+1<t_v$ and $h'+i+1>t_v+1$, which again corresponds to Case II of \Cref{item:each-node-false-gen}. Consider the node $u$ with $\dist(v,u)=t_v-i-1$ and $\dist(u,w)=i-(t_v-d_v)+1$ on that path. Let $u'$ be the neighbor of $u$ on $\texttt{path}(v,w)$ with $\dist(u',v)=t_v-i-2$ and $\dist(u',w)=i-(t_v-d_v)+2$. We need to show, by the end of iteration $i+1$, there is a tree rooted at $v$ in the identifier-induced graph. Moreover, for all nodes along the length-$(t_v-i-2)$ directed path from $u'$ to $v$ in the identifier-induced graph, their $\flagnew$ value is $false$.
	To see this, first notice that, by the end of iteration $i$, by Case II in \Cref{item:root-tree} of the induction hypothesis, there is a tree rooted at $v$ in the identifier-induced graph, call this tree $T_{v,i}$; moreover, for all nodes along the length-$(t_v-i-1)$ directed path from $u$ to $v$ in $T_{v,i}$, their $\flagnew$ value is $false$.
	Now, in iteration $i+1$, $v$ will not change parent, implying there is still a tree rooted at $v$ by the end of iteration $i+1$. Moreover, for each node $w'$ on the length-$(t_v-i-2)$ directed path from $u'$ to $v$ in $T_{v,i}$, it holds that \highlight{$w'$ will not change parent in iteration $i+1$} and $\flagnew_{w'}=false$ by the end of iteration $i+1$ since at least one child (particularly, its child on the path) has passed a $\flagnew$ with $false$ value to it in iteration $i+1$.

	\item Scenario V: $i+1\geq t_v$. By \Cref{item:each-node-set-rid-sent-gen}, node $v$ sets $\rid_v$ to $f(v,i+1,g_{\min}(v,i+1))\neq\vid_v$ at the end of iteration $i+1$. Thus, it must have a parent at the end of iteration $i+1$.
\end{itemize}

Lastly, we prove \Cref{item:no-termination-gen}.
Since no node sets $\build$ to $false$ at the end of iteration $i$, every node will not receive a termination signal from its neighbors in iteration $i+1$. Thus, at the end of iteration $i+1$, any node with a parent will not set $\build$ to $false$ by itself. For a node $v$ without a parent, according to \Cref{item:each-node-false-gen} which is already proved for iteration $i+1$, it has  $\flagnew_v=false$ at the end of iteration $i+1$. Thus, a node without a parent will not set $\build$ to $false$ at the end of iteration $i+1$. This completes the proof of the inductive step.

At this point, the lemma can be easily proved. By \Cref{item:no-termination-gen}, no node quits the BFS-tree building procedure within $d+\rounds-1$ iterations. By \Cref{item:each-node-set-rid}, at the end of iteration $d+\rounds-1$, all nodes have identical $\rid$, which is $\vid_{v_{min}}$. By Case I of \Cref{item:each-node-false-gen}, at the end of iteration \highlight{$d+Q-1$}, all nodes form a single tree rooted at $v_{min}$ in the identifier-induced graph. Moreover, no node will ever change parent or $\rid$ later.
\end{proof}

\begin{proof}[\underline{Proof of \Cref{lem:comp-min-subtree-D-gen}}]
By algorithm description, $v$ will never choose a neighbor as parent since it holds a global minimum token, thus $v$ is always a root in the identifier-induced graph.

Consider an arbitrary iteration $i$, assume node $u$ is in the identifier-induced subtree rooted at $v$ by the end of iteration $i$. Consider the path from $u$ to $v$ in this identifier-induced subtree, we claim this path must be a shortest path connecting $u$ and $v$ in the network graph $G$. Otherwise, it means that on every shortest path connecting $u$ and $v$ in the network graph $G$, there is a node $w$ that quits the BFS-tree building procedure (that is, sets $\build_w=false$) before iteration $i_w$, where $i_w$ represents the iteration in which $w$ should receive the last piece of $\rid_v$, which equals the the distance between $w$ and $v$ in $G$ plus $O(\frac{L}{\log(L/B)})$. But in such case, due to the flooding mechanism of a $false$-valued $\build$ variable, before $u$ can receive the last piece of $\rid_v$, it will set $\build_u$ to $false$, hence will not join the identifier-induced subtree rooted at $v$, resulting in a contradiction. Hence, the path from $u$ to $v$ in the identifier-induced subtree rooted at $v$ must be a shortest path connecting $u$ and $v$ in the network graph $G$. Therefore, the length of this path is $O(D)$.
\end{proof}

\begin{proof}[\underline{Proof of \Cref{lem:comp-all-quit-BFS-gen}}]
To prove the lemma, we only need to show that some node will quit \textsc{BFS-Tree-Building} within $O(D + \frac{L}{\log(L/B)}) $ iterations. This is because the flooding mechanism of a $false$-valued $\build$ variable ensures, once a node $v$ sets $\build_v=false$, all other nodes will set $\build$ to $false$ within (at most) another $D$ iterations.

If some node quits \textsc{BFS-Tree-Building} before the global minimum token value is known by every node, then we are done. Hence, assume that this is not true. Then, by the end of iteration ${i} =  O(D + \frac{L}{\log(L/B)}) $, global minimum token value is known by every node. Particularly, each node has an $\rid$ with a value equal to the global minimum token. Within another $O(\frac{L}{\log(L/B)})$ iterations, every node has sent its complete identifier to neighbors (that is, $\sent = \ceillb$). At that point, each node is in some identifier-induced subtree rooted at some node that has a global minimum token as input. Moreover, no node will change its $\rid$ or parent ever since. By \Cref{lem:comp-min-subtree-D-gen}, any tree rooted at some node that has a global minimum token as input has $O(D)$ height.

From that point onward, by our algorithm, nodes within any such tree will start setting $\flagnew$ to $true$ from leaves to root. Since the height of any such tree is $O(D)$, after $O(D)$ iterations, either some node already sets $\build$ to $false$ and quits \textsc{BFS-Tree-Building}, or some root of such tree sets $\build$ to $false$ and quits \textsc{BFS-Tree-Building}. In both cases, some node quits \textsc{BFS-Tree-Building} within $O(D + \frac{L}{\log(L/B)})$ iterations since the start of execution. 
\end{proof}

\begin{proof}[\underline{Proof of \Cref{lem:comp-all-quit-decide-gen}}]
Recall that our algorithm guarantees if one node generates an $\res\neq\perp$ and then halts, then this $\res$ is broadcast to all other nodes. Hence, to prove the lemma, we show some node will halt within $O(D+k \cdot \frac{L}{\log(L/B)})$ iterations (recall that each iteration consists of $\Theta(\frac{\log(L/B)}{B})$ rounds). To this end, we will show that after all nodes quit \textsc{BFS-Tree-Building} which happens within $O(D + \frac{L}{\log(L/B)})$ iterations (by \Cref{lem:comp-all-quit-BFS-gen}), there exists a tree in the identifier-induced graph of height $O(D)$ (by \Cref{lem:comp-min-subtree-D-gen}), and convergecast tokens inside this tree takes $O(D+k \cdot \frac{L}{\log(L/B)})$ iterations.

Recall that in our algorithm, after all nodes have sent $\build=false$ and $\sent = \ceillb$ to neighbors, and before any node halts, a node is always sending tokens to its parent if its token list is not empty. We argue that, after all nodes have sent $\build = false$ and $\sent = \ceillb$ to neighbors, if an identifier-induced subtree rooted at node $v$ has height $h_v$ and contains $k_v$ tokens, then $v$ can collect each of the $k_v$ tokens exactly once within $O(h_v+k_v \cdot \frac{L}{\log(L/B)})$ iterations. To see this, notice that in our algorithm a token is divided into $\ceillb$ pieces and $\Theta(\frac{\log{(L/B)}}{B})$ such pieces are sent in one iteration, so it takes $O(\frac{L}{\log(L/B)} )$ iterations for a node to send a complete token.

We can adopt our previous result in the proof of \Cref{lem:comp-all-quit-decide}. In particular, notice that collecting $O(k_v)$ large tokens from an $h_v$-height identifier-induced subtree in $T$ iterations is equivalent to collecting $O(k_v \cdot \frac{L}{\log(L/B)})$ small tokens from $h_v$-height identifier-induced subtree in $T$ rounds. By the analysis in the proof of \Cref{lem:comp-all-quit-decide}, we know $T=O(h_v + k_v \cdot \frac{L}{\log(L/B)})$.

By now, we have proved that, after all nodes have sent $\build=false$ and $\sent = \ceillb$ to neighbors, if an identifier-induced subtree rooted at $v$ has height $h_v$ and contains $k_v$ tokens, then $v$ can collect each token exactly once in the tree in $O(h_v + k_v \cdot \frac{L}{\log(L/B)})$ iterations.

By \Cref{lem:comp-all-quit-BFS-gen}, we know that all nodes have sent $\build = false$ to neighbors in $O(D + \frac{L}{\log(L/B)})$ \highlightblue{iterations}.
Then, consider a node $v$ which has a global minimum token as input. In the identifier-induced graph, by \Cref{lem:comp-min-subtree-D-gen}, there is a tree rooted at $v$ and the identifier-induced subtree rooted at $v$ has height $O(D)$ and contains at most $k$ tokens. By our above analysis, in $O(D + k \cdot \frac{L}{\log(L/B)})$ iterations, either $v$ collects all tokens in \highlight{this identifier-induced subtree}, or some node already halts. Moreover, it is easy to see in \highlight{$O(D)$} iterations, either $v$ has obtained a count on the size of \highlight{this identifier-induced subtree} (that is, $\cnt\neq\perp$), or some node already halts.

Recall that one iteration consists of \highlight{$\Theta(\frac{\log(L/B)}{B})$} rounds, we conclude, within \highlightblue{ \[O\left(D \cdot \frac{\log(L/B)}{B}  +k \cdot \frac{L}{B} \right)=O\left(D \cdot \frac{\log(L/\log n)}{\log n}  +k \cdot \frac{L}{\log n} \right)\]} rounds since the start of execution, some node will halt.
\end{proof}

\end{document}